\documentclass[usenatbib]{mn2e}
\usepackage{times}
\usepackage{txfonts}
\usepackage{epsfig}
\usepackage{graphicx}
\usepackage[margin=0.5in, top=1in, bottom=1in, a4paper, pdftex, dvips]{geometry}

\title[Quasi-biennial variations in p-mode frequencies] {Quasi-Biennial variations in helioseismic frequencies: Can the source of the variation be localized?}
\author[A.-M. Broomhall et al.]{A.-M.
Broomhall$^1$\thanks{amb@bison.ph.bham.ac.uk}, W.~J. Chaplin$^1$, Y.
Elsworth$^1$, R. Simoniello$^2$\\ $^1$School of Physics and
Astronomy,
University of Birmingham, Edgbaston, Birmingham B15 2TT\\
$^2$PMOD/WRC Physikalisch-Meteorologisches Observatorium Davos-World
Radiation Center, 7260 Davos Dorf, Switzerland}

\begin{document}
\maketitle \begin{abstract} We investigate the spherical harmonic
degree ($l$) dependence of the ``seismic'' quasi-biennial
oscillation (QBO) observed in low-degree solar p-mode frequencies, using
Sun-as-a-star Birmingham Solar Oscillations Network (BiSON) data. The amplitude of the
seismic QBO is modulated by the 11-yr solar cycle, with the amplitude
of the signal being largest at solar maximum. The amplitude of the signal is
noticeably larger for the $l=2$ and 3 modes than for the $l=0$ and 1
modes. The seismic QBO shows some frequency dependence but this dependence is not as strong as observed in the 11-yr solar cycle. These results are consistent with the seismic
QBO having its origins in shallow layers of the interior (one possibility being the bottom of the shear
layer extending 5\,per cent below the solar surface). Under this scenario the magnetic
flux responsible for the seismic QBO is brought to the surface (where its
influence on the p modes is stronger) by buoyant flux from the 11-yr cycle, the strong component of which is observed at
predominantly low-latitudes. As the $l=2$ and 3 modes are much more
sensitive to equatorial latitudes than the $l=0$ and 1 modes the
influence of the 11-yr cycle on the seismic QBO is more visible in $l=2$ and 3
mode frequencies. Our results imply that close to solar maximum the main influence of the seismic QBO occurs at
low latitudes $(<45^\circ)$, which is where the strong component of
the 11-yr solar cycle resides. To isolate the latitudinal dependence of the seismic QBO from the 11-yr solar cycle we must consider epochs when the 11-yr solar cycle is weak. However, away from solar maximum, the amplitude of the seismic QBO is weak making the latitudinal dependence hard to constrain.
\end{abstract}

\section{Introduction}

It has been known since the mid 1980s that the frequencies of the
Sun's resonant oscillations, known as p modes, vary throughout the
solar cycle with the frequencies being at their highest when the
solar activity is at its maximum \citep[e.g.][]{Woodard1985}. More
recently a quasi-biennial (2-year) signal has been observed in the
frequencies of low-degree solar p modes \citep[e.g.][and references
therein]{Fletcher2010}. In this paper we investigate the same
quasi-biennial signal in more detail. The seismic quasi-biennial
signal that we observe is highly correlated with the
``quasi-biennial oscillation'' (QBO) associated in
other proxies of the Sun's activity \citep[e.g.][]{Kane2005, Li2006, Ivanov2007, Vecchio2009, Zaqarashvili2010, Badalyan2011}, and here we refer to the
quasi-biennial signal observed in the p-mode frequencies as the
``seismic'' QBO. We note here that the QBO observed in other proxies
of the Sun's activity has previously been used to refer to signals
with periodicities ranging from 1-3\,yrs. The actual period varies
depending on which epoch is considered and which proxy for solar
activity is used \citep[e.g.][and references therein]{Kane2005}. There are many different methods by which a quasi-biennial modulation of the 11-yr solar cycle could be generated \citep[e.g.][and references therein]{Hoyng1990, Benevolenskaya1998a, Benevolenskaya1998b, Tobias2002, Wang2003, Zaqarashvili2010}, some of which involve the near-surface rotational shear layer.

The horizontal spatial structure of p modes in the Sun can be
described by spherical harmonics with each mode being characterized
by its spherical harmonic degree $(l)$ and azimuthal order $(m)$.
The spatial structure of the oscillations means that
the sectoral $|m|=l$ components of a mode are more concentrated
around the equator than the zonal components, and this is
particularly true as $l$ increases \citep[see, for
example,][]{JCD2002}. The sectoral components are predominantly
sensitive to changes in activity at lower latitudes. This is where
strong-field structures, such as sunspots, dominate. On the other
hand the zonal ($m=0$) components show a greater relative sensitivity
to higher latitudes, where only the weak-component flux is present.
Although the weak-component flux does increase slightly with the
solar activity cycle it is the substantial increase in the
strong-field component at low latitudes
\citep[$\le45^{\circ}$;][]{DeToma2000} that is largely responsible
for the observed p-mode solar cycle variations \citep{Howe1999}.
Therefore, the sectoral components of a p mode are more sensitive to
solar cycle perturbations than the zonal components. The sizes of the shifts of different mode components
scale proportionately with the corresponding spherical harmonic
components of the observed line-of-sight surface magnetic field
\citep[e.g.][]{Howe1999, Chaplin2003, Chaplin2004b, Chaplin2004,
Jimenez2004, Chaplin2007}. By examining the $l$-dependence of the
observed changes in frequencies throughout the solar cycle we can
gain information on the latitudinal dependence of the solar magnetic
field. Furthermore, the latitudinal dependence of the 11-yr solar cycle can be constrained using the frequencies of low-$l$ modes \citep{Chaplin2007a, Chaplin2011}.

Here we investigate whether the same constraints can be put on the latitudinal dependence of the seismic QBO, using low-$l$ modes, by
examining the frequency shifts observed in modes of different $l$. \citet{Vecchio2009} used the green coronal emission line at
530.3\,nm observed between 1939 and 2005 to examine the spatial
distribution of the QBO. They found that the signal was
strong in polar regions and weaker around the equator.

We make use of Sun-as-a-star (Doppler velocity) observations, which are sensitive to p modes with the largest horizontal scales (or lowest $l$). Resolved helioseismic observations allow higher-$l$ modes to be observed. However, in general these modes are more localized than are the low-$l$ modes and hence sensitive to local conditions. This has the potential to confuse the discussion of patterns on a large scale, which is what interests us here. Furthermore, the method used in the determination of frequencies for the high-$l$ modes is different from that used at low $l$ because of the influence of mode leakage, which occurs because the spatial filters are not perfect due to only half of the Sun being observed. Although analysis based on combined data sets is done, great care has to be taken to minimize systematic discrepancies \citep[e.g.][]{Chaplin2004b, Chaplin2004a}. Therefore, in this paper, we assess what information concerning the location of the QBO can be obtained from low-$l$ modes alone.

The structure of the remainder of this paper is as follows: Section
\ref{section[data]} describes the data used in this paper. The
$l$-dependence of the raw frequency shifts is discussed in Section
\ref{section[freq shifts]}. In Section \ref{subsection[significance
of QBO]} we assess whether the observed seismic QBO is strong enough to be significant when the frequencies are examined for each $l$ individually.
Then, in Section \ref{section[QBS]}, the $l$-dependence and frequency dependence of the
seismic QBO is discussed. In Section \ref{section[models]} we use
simple models to try and constrain the latitudinal dependence of the
seismic QBO. Finally the main results are summarized in Section
\ref{section[discussion]}.

\section{Data}\label{section[data]}

The Birmingham Solar-Oscillations Network (BiSON) makes
Sun-as-a-star (unresolved) Doppler velocity observations. We have
used the BiSON data to investigate whether a seismic QBO is present
in the frequencies of modes with $0\le l\le 3$, i.e. those modes
that are prominent enough in Sun-as-a-star data to allow their
frequencies to be determined with enough precision and accuracy to
study solar cycle variations.

BiSON has now been collecting data for over 30 yrs. The time
coverage of the early data, however, is poor compared to more recent
data. As a consequence, here we have chosen to analyse the mode frequencies observed by BiSON
from 1992 October 10 to 2011 April 7. To study the variation in the
p-mode frequencies with time, the observations made by BiSON were
divided into both 182.5-day-long independent subsets and
365-day-long subsets that overlapped by 91.25\,d. The time series used in this paper cover a shorter epoch than the work of \citet{Fletcher2010}. The start date for
this analysis, i.e. 1992 October 10, was chosen because after this
date the duty cycles of the 182.5-d time series were always above
60\,per cent, and the average duty cycle of the subsets was
80\,per cent. A high duty cycle was particularly important for
determining accurately the frequencies of $l=3$ (and to a lesser
extent the $l=2$) modes. This is because the heights of the
individual $m$ components of these modes in a frequency-power
spectrum constructed from Sun-as-a-star data are less prominent than
the individual $m$ components of the $l=0$ and 1 modes. We have also classified the subsets as occurring during periods of either high- or low-surface activity. We have
defined the level of surface activity to be ``high'' when the mean
10.7\,cm flux observed in the 182.5\,d independent subsets is above
$100\times10^{-22}\,\rm W\,m^{-2}\,Hz^{-1}$.

Estimates of the mode frequencies were extracted from each subset by
fitting a modified Lorentzian model to the data using a standard
likelihood maximization method \citep{Fletcher2009}. A reference
frequency set was determined by averaging the frequencies in subsets
covering the minimum activity epoch at the boundary between cycle 22
and cycle 23. It should be noted that the main results of this paper
are insensitive to the exact choice of subsets used to make the
reference frequency set. Frequency shifts were then defined as the
differences between frequencies given in the reference set and the
frequencies of the corresponding modes observed at different epochs
in other subsets \citep{Broomhall2009}.

For each subset in time and each $l$ in the range $0\le l\le 3$,
three weighted-average frequency shifts were generated, where the
weights were determined by the formal errors on the fitted
frequencies: first, a ``total'' average shift was determined by
averaging the individual shifts of the modes over fourteen overtones
(covering a frequency range of $1.86-3.72\,\rm mHz$); second, a
``low-frequency'' average shift was computed by averaging over seven
overtones whose frequencies ranged from 1.86 to $2.77\,\rm mHz$; and
third, a ``high-frequency'' average shift was calculated using seven
overtones whose frequencies ranged from 2.80 to $3.72\,\rm mHz$. The
lower limit of this frequency range (i.e., $1.86\,\rm mHz$) was
determined by how low in frequency it was possible to accurately fit
the data before the modes were no longer prominent above the
background noise. The upper limit on the frequency range (i.e.,
$3.72\,\rm mHz$) was determined by how high in frequency the data
could be fitted before errors on the obtained frequencies became too
large to obtain precise values for the frequency shifts due to
increasing line widths causing modes to overlap in frequency.

\section{Degree dependence of frequency shifts over 11-yr solar cycle}\label{section[freq
shifts]}

\begin{figure*}
  \centering
  \includegraphics[width=0.48\textwidth, clip]{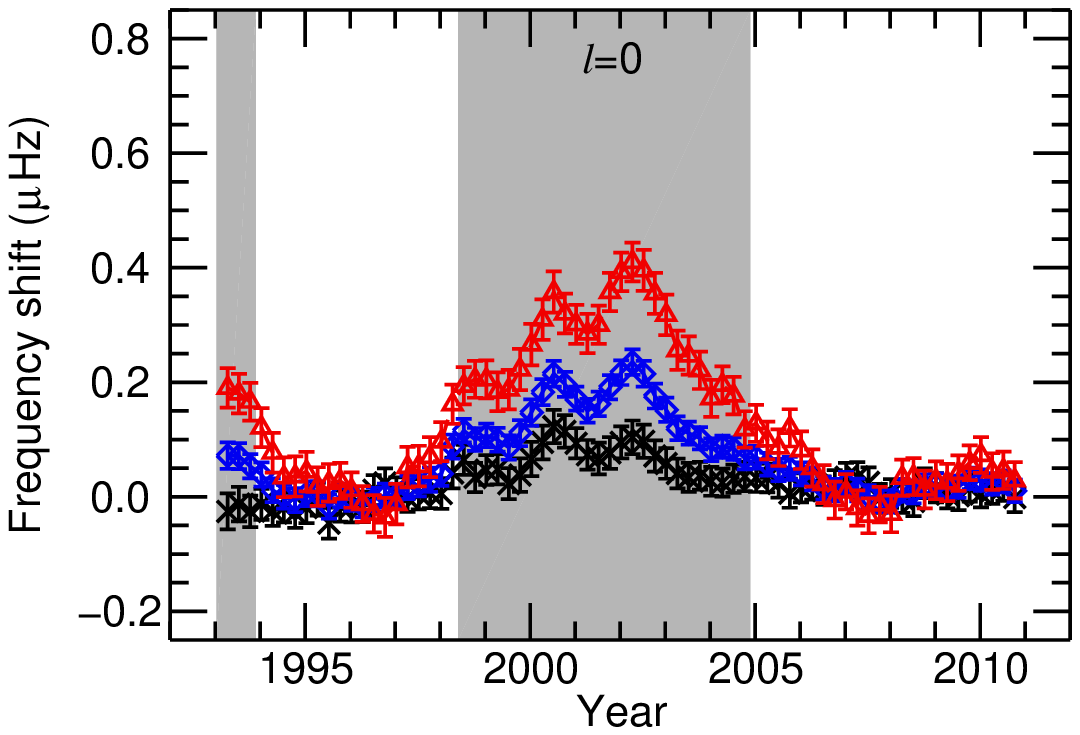}
  \includegraphics[width=0.48\textwidth, clip]{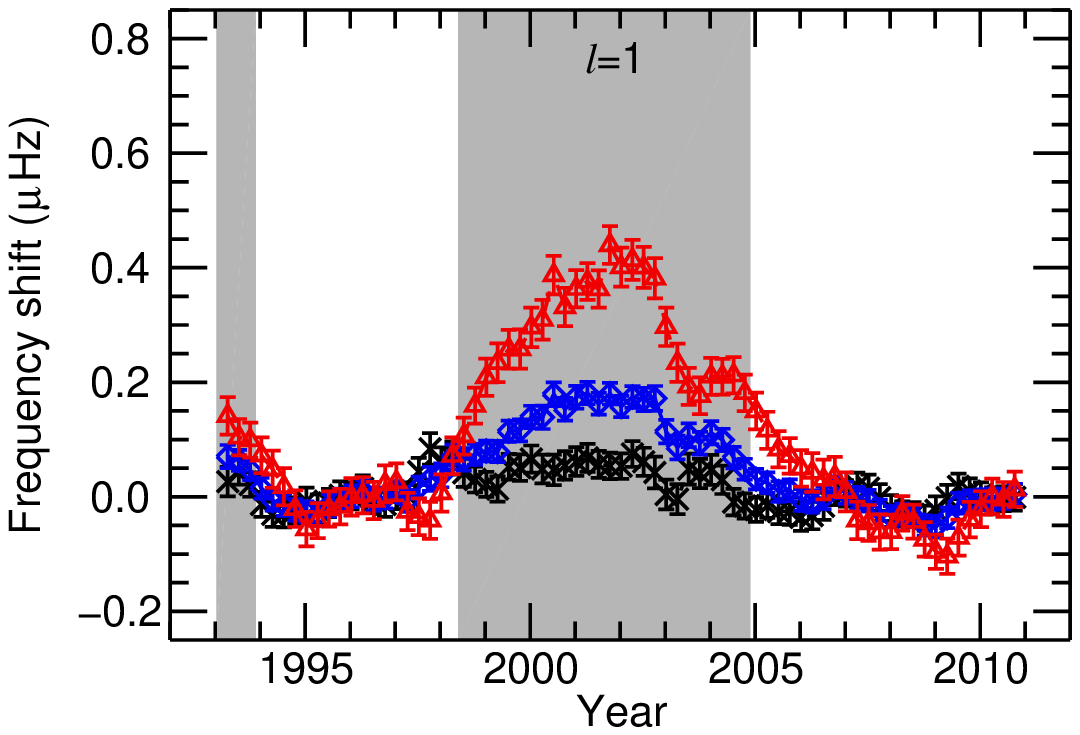}\\
  \includegraphics[width=0.48\textwidth, clip]{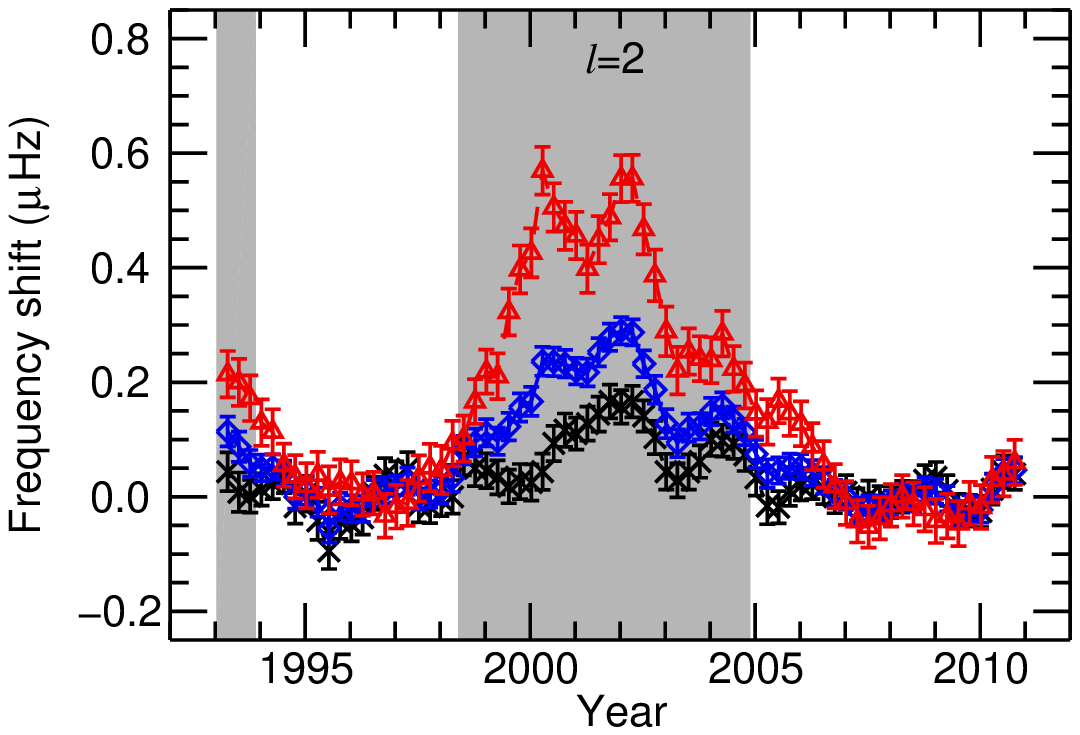}
  \includegraphics[width=0.48\textwidth, clip]{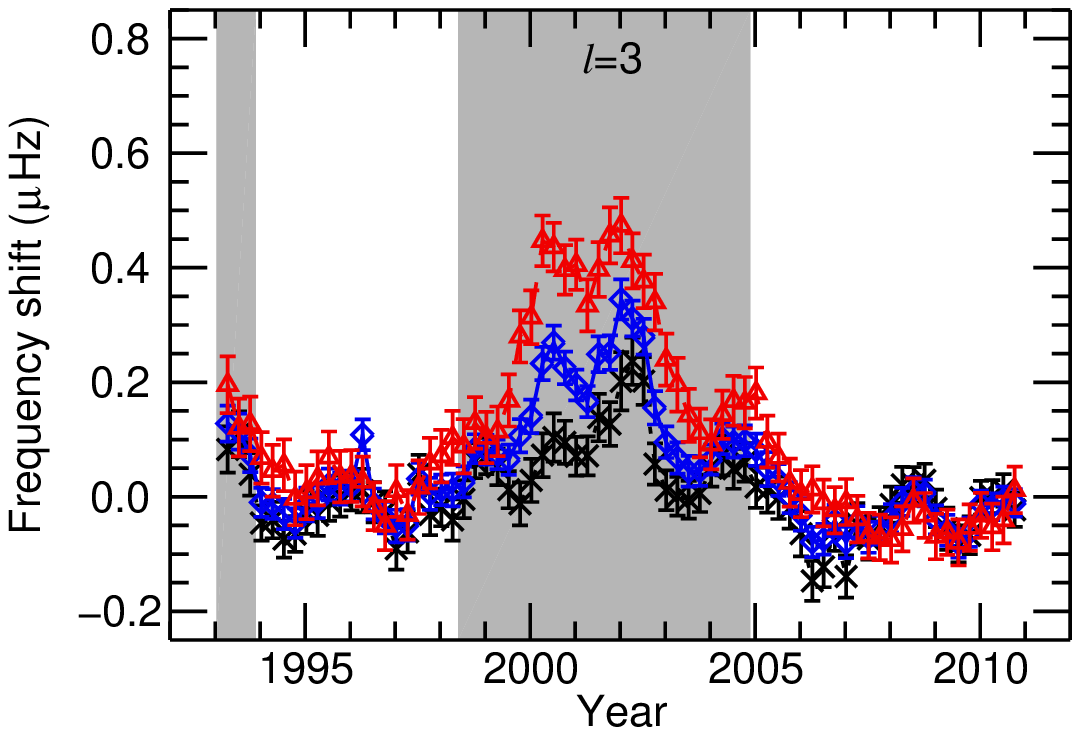}\\
  \caption{Individual-$l$ average frequency shifts of ``Sun-as-a-star'' modes observed in 365\,d subsets that overlapped by 91.25\,d, with
  frequencies between 1.86 and $3.72\,\rm mHz$ (total-frequency band, blue solid line, and diamond symbols); 1.86 and $2.77\,\rm
  mHz$ (low-frequency band, black dotted line, and cross symbols); and 2.80 and $3.72\,\rm mHz$
  (high-frequency band, red dashed line, and triangle symbols). In each panel the shaded regions
  denote times of high-surface activity,
  while the unshaded regions denote times of
  low-surface activity.}\label{figure[freq shifts]}
\end{figure*}

Fig. \ref{figure[freq shifts]} shows the individual-$l$ averaged
frequency shifts observed in the 365\,d subsets that overlapped by 91.25\,d. The
results from the 182.5\,d subsets and the 365\,d subsets are
consistent and so we have chosen to plot the 365\,d subsets
only but we note that the 182.5\,d independent subsets were
used to assess the statistical significance of any signals in the
frequency shifts as the overlaps in the 365\,d subsets introduce
artifacts in the periodicity measures we used for this purpose. The shaded regions
in Fig. \ref{figure[freq shifts]} differentiate between times of
high- (shaded) and low-surface activity (unshaded).

The 11-yr solar cycle is clearly visible in the frequency shifts.
For each $l$ the high-frequency range shows more sensitivity to the
11-yr cycle than the low-frequency range. This is consistent with
previous observations \cite[e.g.][and references
therein]{Broomhall2009}.

The higher-$l$ modes are more sensitive to the 11-yr solar cycle
than the lower-$l$ modes. For example, the $l=2$ and 3 modes show a
larger frequency shift at times of high activity than the $l=0$ and
1 modes. This is in agreement with previous observations and, as we will now explain, there are two main reasons for the observed $l$ dependence \citep[e.g.][]{Libbrecht1990, Chaplin1998, Jimenez2001, Salabert2009}. Firstly, at fixed frequency, the mode inertia decreases with increasing degree \citep[e.g.][]{Howe1999, Chaplin2001,  Rabello2008}. The mode inertia \citep{JCD1991} is a measure of the interior mass affected by a given mode and so, as the mode inertia decreases a smaller volume of the interior is associated with the motions generated by the mode. As such the frequencies of high-$l$ modes are more susceptible to change throughout the solar cycle than low-$l$ modes. However, the mode inertia varies very little across the modes examined here (at $3000\,\rm\mu Hz$ the inertia of an $l=0$ mode is approximately 0.3\,per cent greater than the inertia of an $l=3$ mode). The second reason for the $l$-dependence of the solar cycle shifts, which is more important for low-$l$ modes than the mode inertia, is the latitudinal distribution of the solar magnetic field at the surface \citep[e.g.][]{Howe1999, Chaplin2003, Chaplin2004b, Chaplin2004, Jimenez2004, Chaplin2007}. Sun-as-a-star observations are most sensitive to the sectoral components of a mode, which, because of the spatial structure of the mode, show a greater sensitivity to equatorial regions as $l$ increases. Therefore, since at solar maximum the solar activity is also concentrated at low latitudes, $l=2$ and 3 modes experience a larger shift than the $l=0$ and 1 modes (when observed in Sun-as-a-star data).

Fig. \ref{figure[freq shifts]} shows that, for each $l$, there is
shorter-term, quasi-biennial structure on top of the dominant 11-yr
trend and we now study this structure in more detail.

\section{Assessing the significance of the seismic QBO}\label{subsection[significance of QBO]}
We begin by assessing whether the quasi-biennial structure observed
in the frequency shifts represent a significant QBO. The
significance of the seismic QBO was assessed by computing
periodograms of the raw frequency shifts. When assessing the statistical
significance of the seismic QBO we used the frequency shifts observed
in the 182.5\,d independent time series as artifacts can be introduced
when using overlapping subsets. Overlaps in the subsets mean that the
periodograms are modulated by a sinc squared function whose first
zero was at $1.0\,\rm yr^{-1}$ i.e. the inverse of the time
difference between independent subsets. This makes it difficult to
determine whether any significant periodicities are present in the
overlapping data, particularly above $0.5\,\rm yr^{-1}$.

\begin{figure*}
  \centering
  \includegraphics[width=0.48\textwidth, clip]{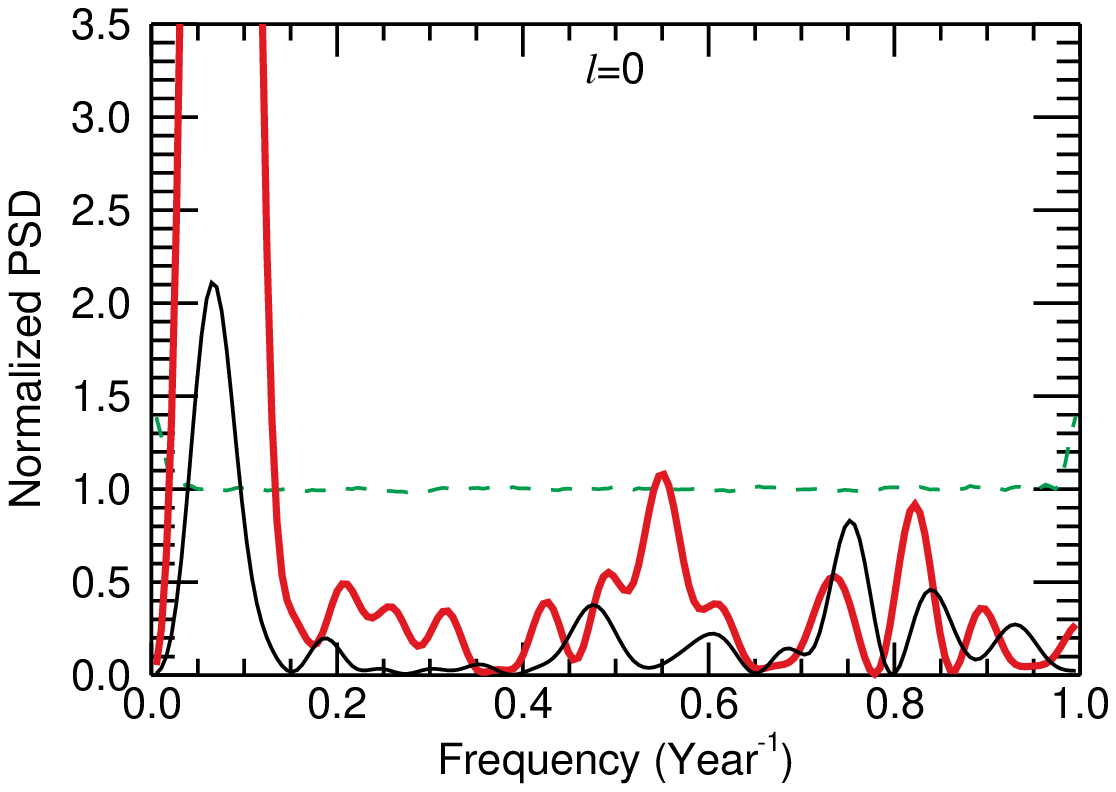}
  \includegraphics[width=0.48\textwidth, clip]{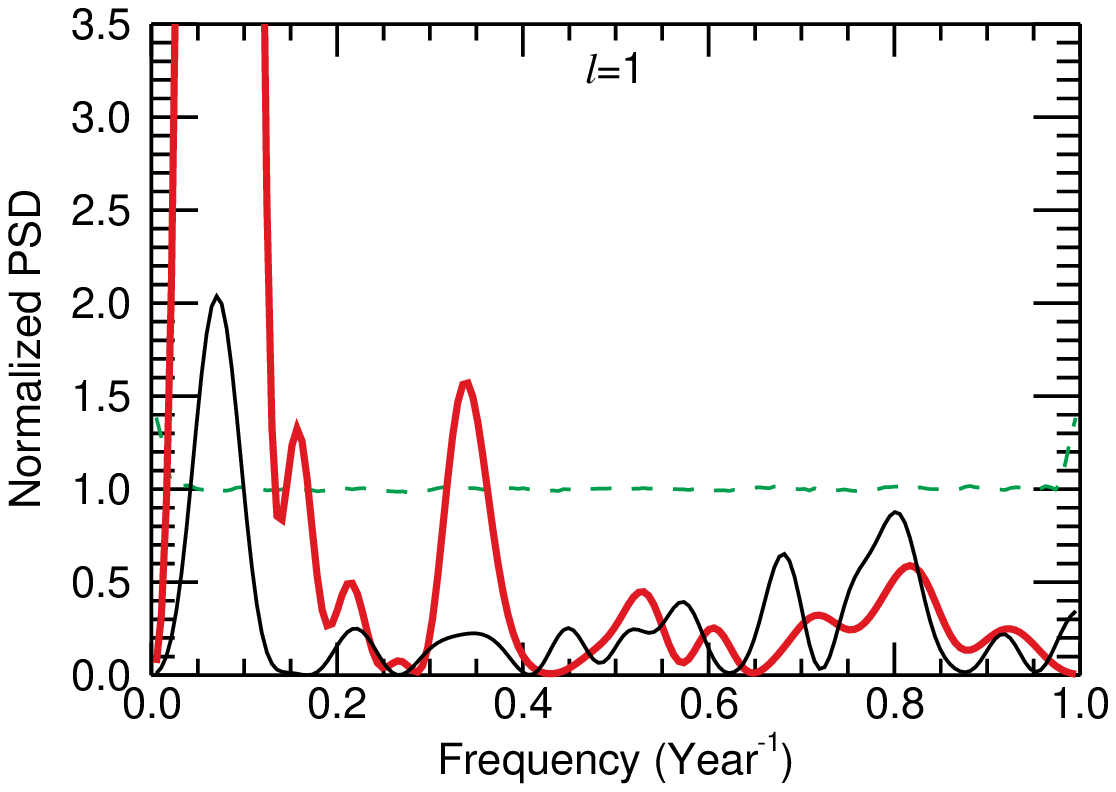}\\
  \includegraphics[width=0.48\textwidth, clip]{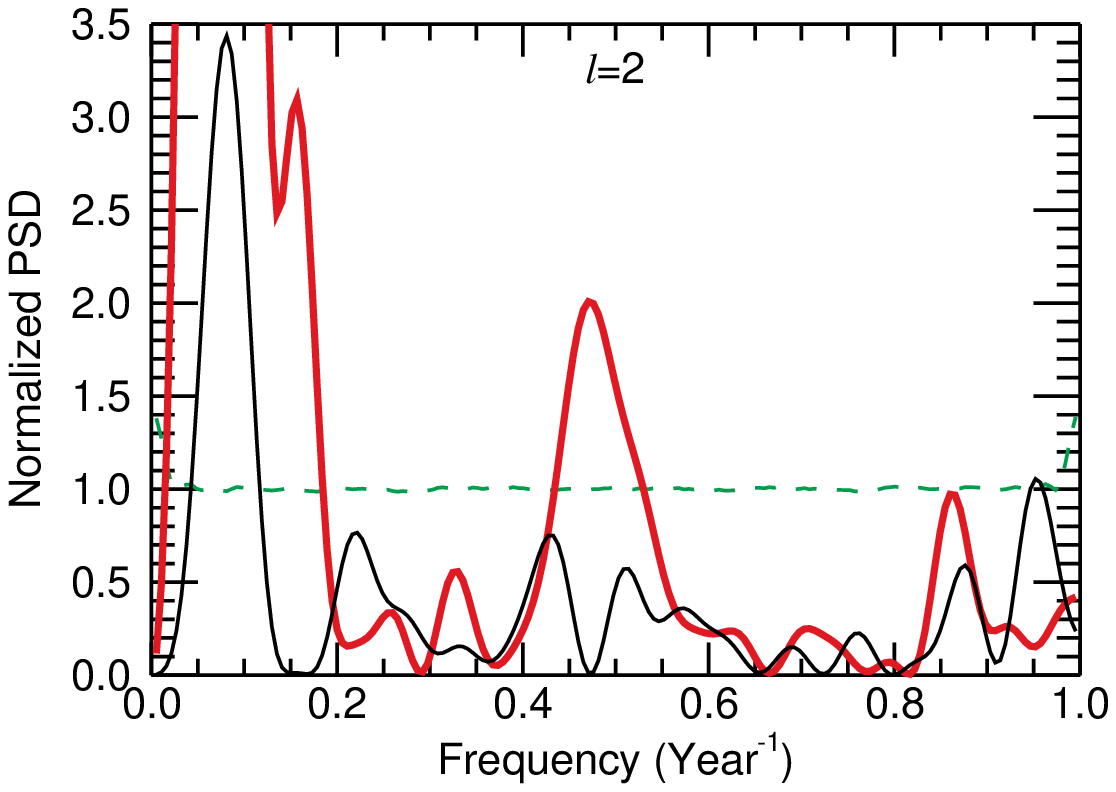}
  \includegraphics[width=0.48\textwidth, clip]{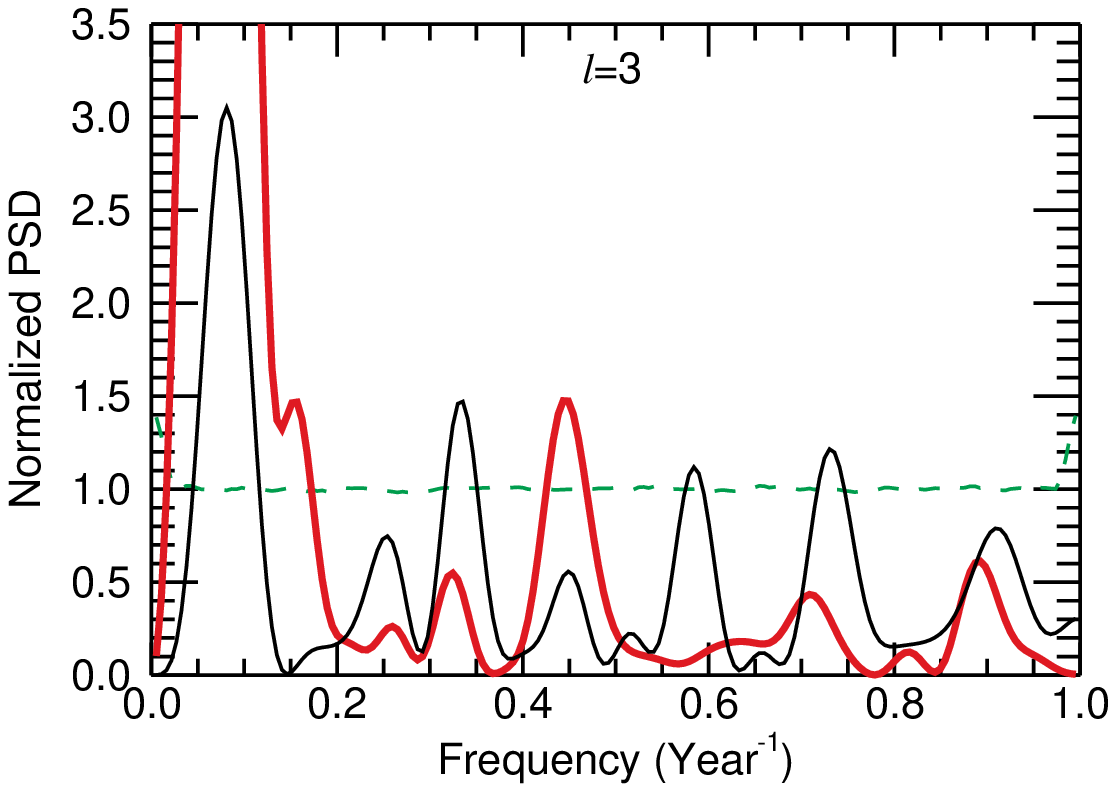}\\
  \caption{Periodograms of the individual-$l$ frequency shifts
  observed in different frequency bands using the 182.5\,d independent subsets. In each panel the black
  thin lines are the periodograms of the low-frequency-range shifts
  and the red thick lines represent the high-frequency-range
  periodograms. The 1\,per
  cent ``false alarm'' significance levels for the respective
  frequency ranges are also plotted (green dashed lines). The
  periodograms have been normalized so that for each periodogram the
  1\,per cent significance levels are unity. Note that the
  underlying resolution of a periodogram of the frequency shifts that had not been
  oversampled is $0.054\,\rm Year^{-1}$.}\label{figure[periodograms]}
\end{figure*}

The mean frequency shift was subtracted
before each periodogram was calculated and the data were oversampled
by a factor of 10. The results are shown in Fig.
\ref{figure[periodograms]}. Also plotted in Fig.
\ref{figure[periodograms]} are the 1\,per cent false alarm
significance levels \citep{Chaplin2002}, which were determined using
Monte Carlo simulations. 100,000 noise time series were simulated, using a normal distribution random number generator, to mimic those plotted in Fig. \ref{figure[freq shifts]}. The standard deviation of each point in the time series was taken as the $1\sigma$ error associated with the rotational splittings plotted in Fig. \ref{figure[freq shifts]}. Periodograms of the time series were calculated and the distribution of the amplitudes observed in the simulated periodograms was used to define the 1\,per cent false alarm significance levels.
The large peaks in each panel of Fig. \ref{figure[periodograms]} at
$0.09\,\rm yr^{-1}$ are the signal from the 11-yr cycle.

The $l=2$ and 3 high-frequency-range periodograms show a significant
peak (at a 1\,per cent false alarm level) at a frequency of
$\sim0.5\,\rm yr^{-1}$. This is the same frequency as the seismic
QBO observed by, for example, \citet{Fletcher2010}. However, we note
that \citeauthor{Fletcher2010} also observed a significant peak at
$\sim0.5\,\rm yr^{-1}$ in low-frequency-range frequency shifts that
were averaged over $0\le l\le2$. However, there is no evidence for a
significant signal at the same frequency in the low-frequency range
observed here. We have improved our analysis techniques since
\citeauthor{Fletcher2010}. However, it is possible that this has introduced some
destructive interference with noise that has reduced the amplitude
of the signal. Alternatively the improved analysis could have
removed constructive interference with noise that was artificially
enhancing the observed signal observed by \citeauthor{Fletcher2010}.
Furthermore, \citeauthor{Fletcher2010} examined a slightly longer
time span, that included two solar maxima, which is when the seismic
QBO is most prominent (this will be shown in Section
\ref{section[QBS]}).

A significant peak is observed at a marginally higher frequency
($\sim0.55\,\rm yr^{-1}$) in the high-frequency-range $l=0$
periodogram. This difference would be equivalent to the resolution
in a non-oversampled periodogram and so may not be significant. The
$l=1$ high-frequency periodogram has a significant peak at a lower
frequency ($\sim0.35\,\rm yr^{-1}$). A significant peak at the same
frequency is also present in the $l=3$ low-frequency-range
periodogram. We recall that the seismic QBO has been used to refer
to oscillations with periodicities ranging from 1-3\,yrs. It is,
therefore, possible that, despite the different periodicities,
the signals observed in the different $l$ modes can be attributed to the same source.

There are two other significant peaks in the $l=3$
low-frequency-range periodogram, at 0.60 and $0.75\,\rm yr^{-1}$
respectively. The origin of these peaks is uncertain. However, we
note that the peak at $0.75\,\rm yr^{-1}$ is of a similar
periodicity to the 1.3-yr signal identified in other studies
\citep[e.g.][]{Howe2000, Jimenez2003, Wang2003, Broomhall2011}.

As a check of our methodology we ran the same tests on some artificial data. The data were simulated for the solar Fitting at Low Angular degree Group \citep[solarFLAG;][]{Chaplin2006}. These data were made in the time domain and were designed to mimic Sun-as-a-star observations. The simulated oscillations were stochastically excited and damped with lifetimes analogous to those of real solar oscillations. The input mode frequencies were the same in each simulated data set and were held stationary throughout each simulated time series. However, the data simulated the stochastic nature of the mode excitation giving access to different mode and noise realizations. We have created a time series whose length was chosen to contain the same number of 182.5\,d subsets as the BiSON data analyzed here. Although not shown here only one significant peak was observed in the periodograms of the frequency shifts. Furthermore, the peak was observed in the $l=3$ frequency shifts and, as we shall explain in Section \ref{section[QBO constraints]}, estimates of the $l=3$ mode frequencies are more noisy than the frequencies estimated for the other $l$. This indicates that we must be careful concerning the significance of the peaks observed in the $l=3$ periodograms as it is possible that the uncertainties associated with the $l=3$ frequencies underestimate the scatter in the estimated frequencies.

\section{The seismic QBO}\label{section[QBS]}

\begin{figure*}
  \centering
  \includegraphics[width=0.33\textwidth,
  clip]{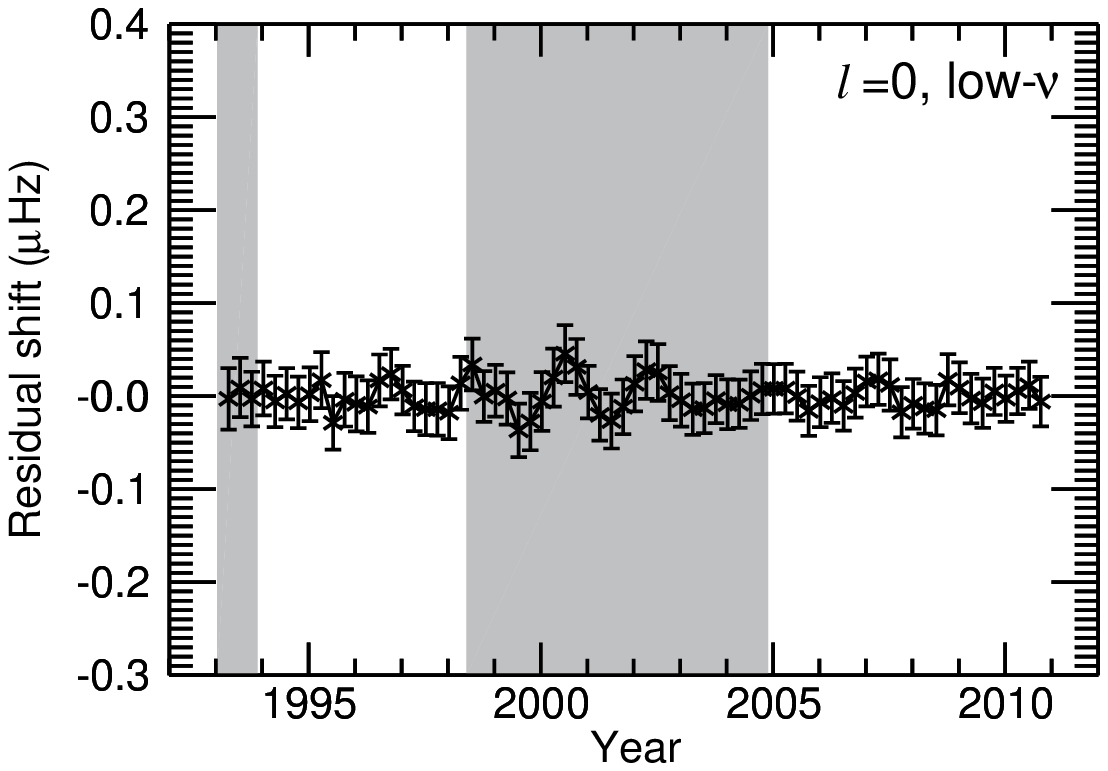}
  \includegraphics[width=0.33\textwidth,
  clip]{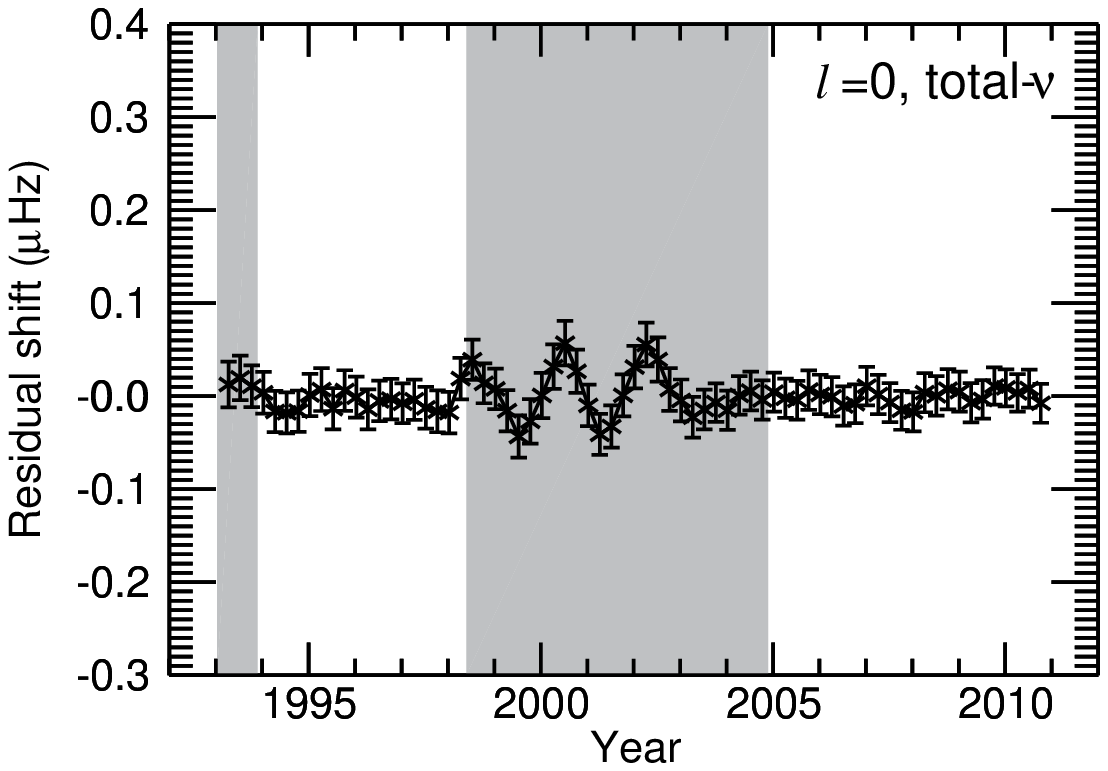}
  \includegraphics[width=0.33\textwidth, clip]{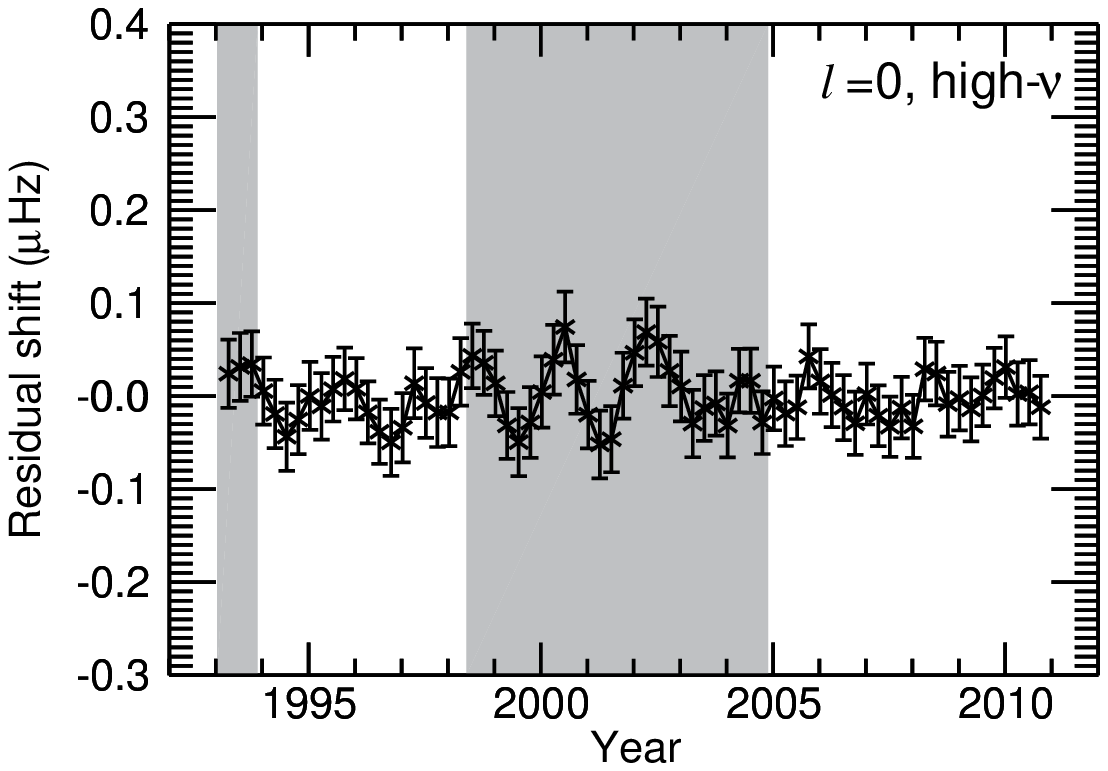}\\
  \vspace{0.5cm}
  \includegraphics[width=0.33\textwidth,
  clip]{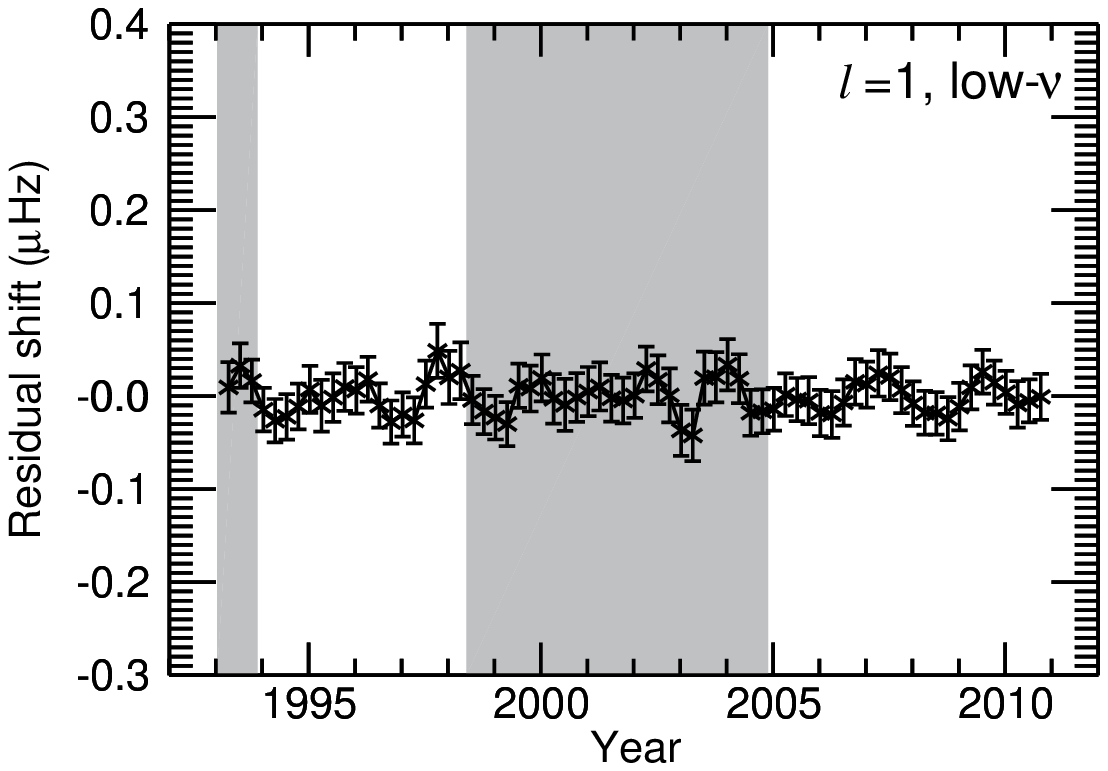}
  \includegraphics[width=0.33\textwidth,
  clip]{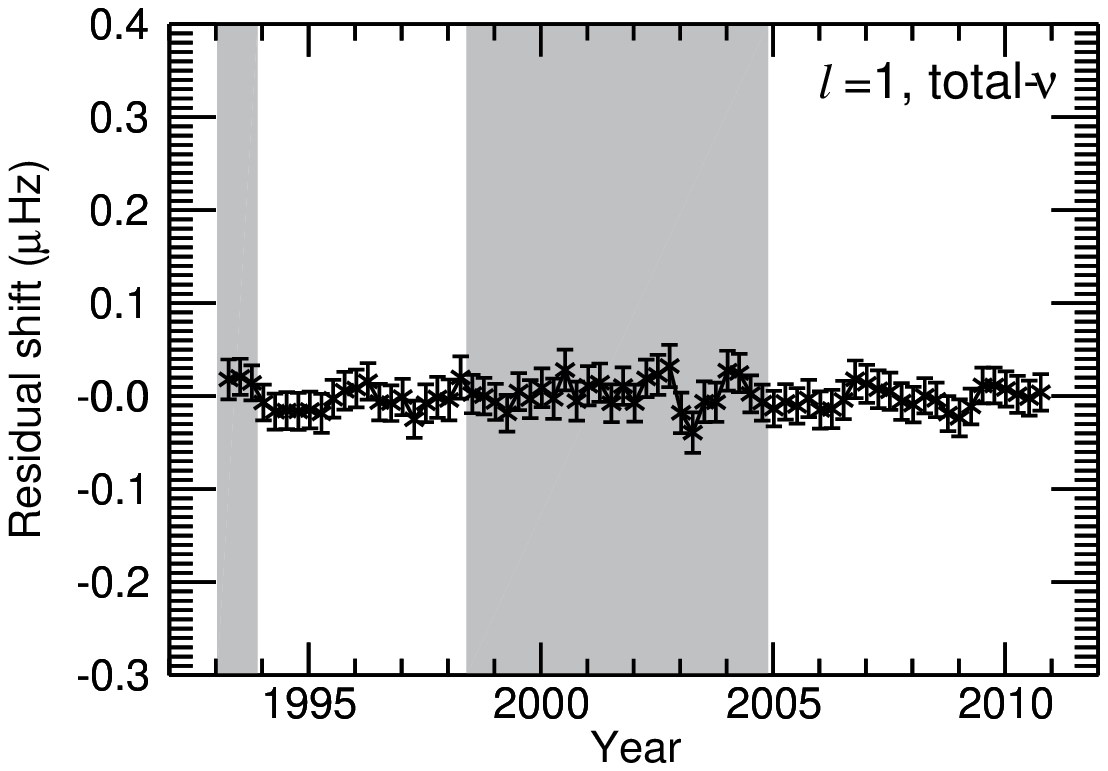}
  \includegraphics[width=0.33\textwidth, clip]{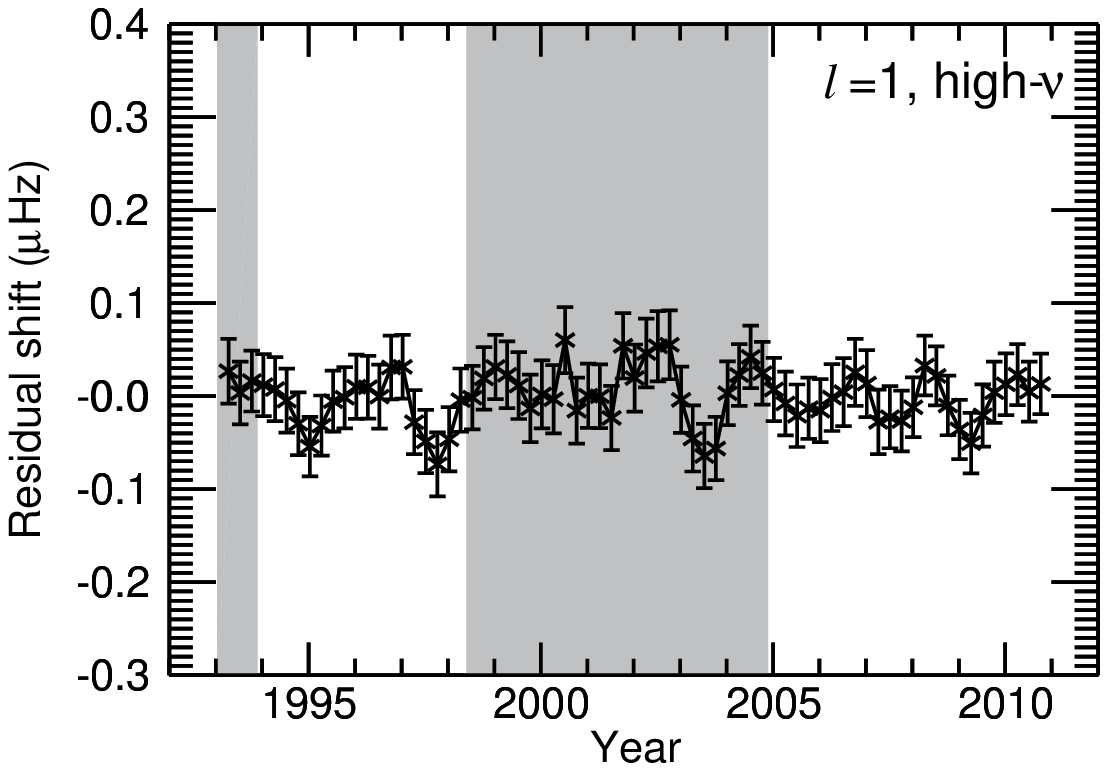}\\
  \vspace{0.5cm}
  \includegraphics[width=0.33\textwidth,
  clip]{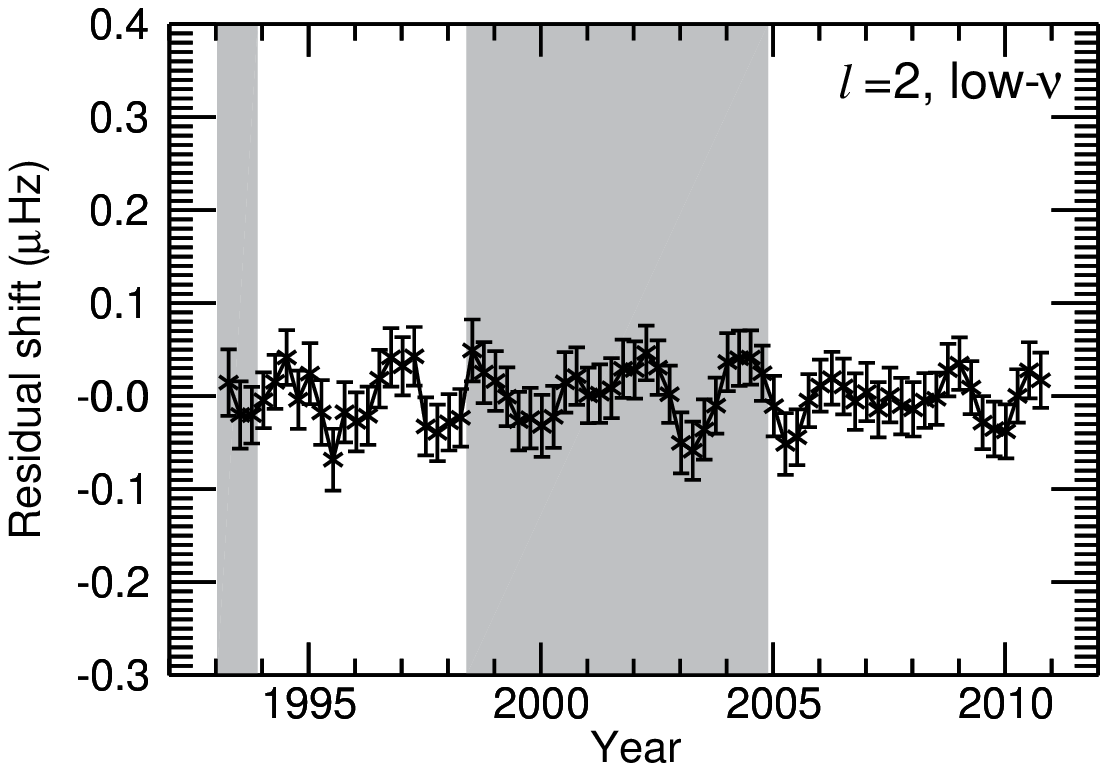}
  \includegraphics[width=0.33\textwidth,
  clip]{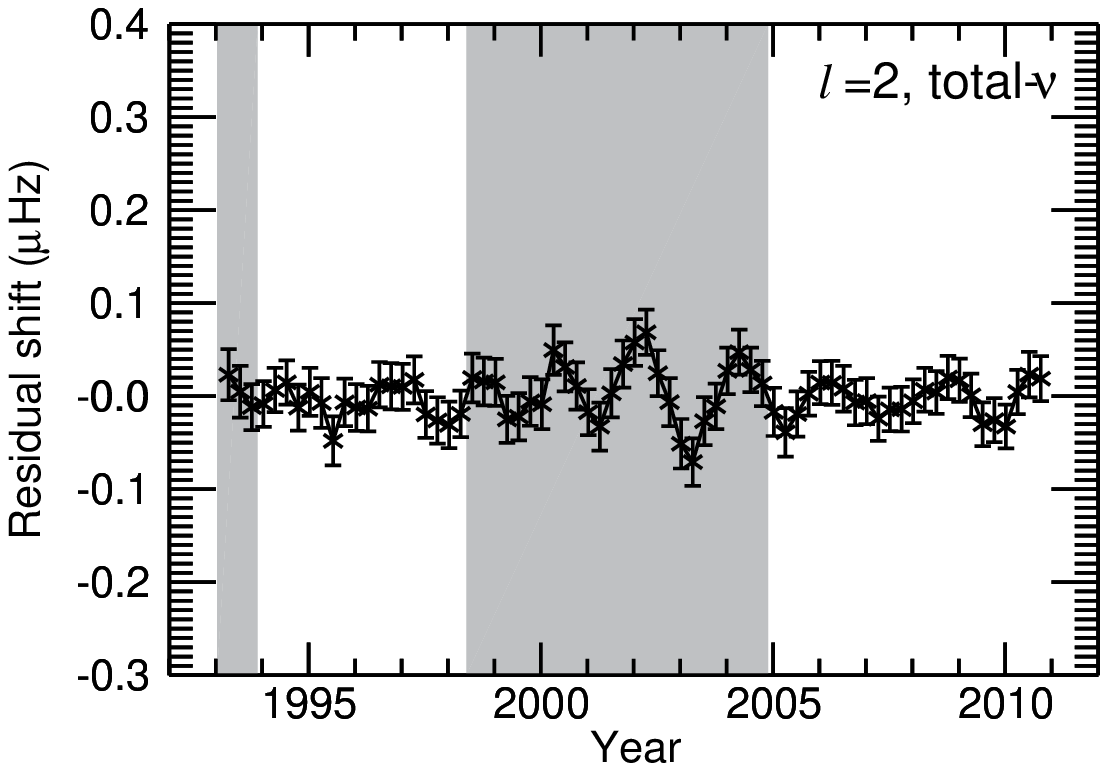}
  \includegraphics[width=0.33\textwidth, clip]{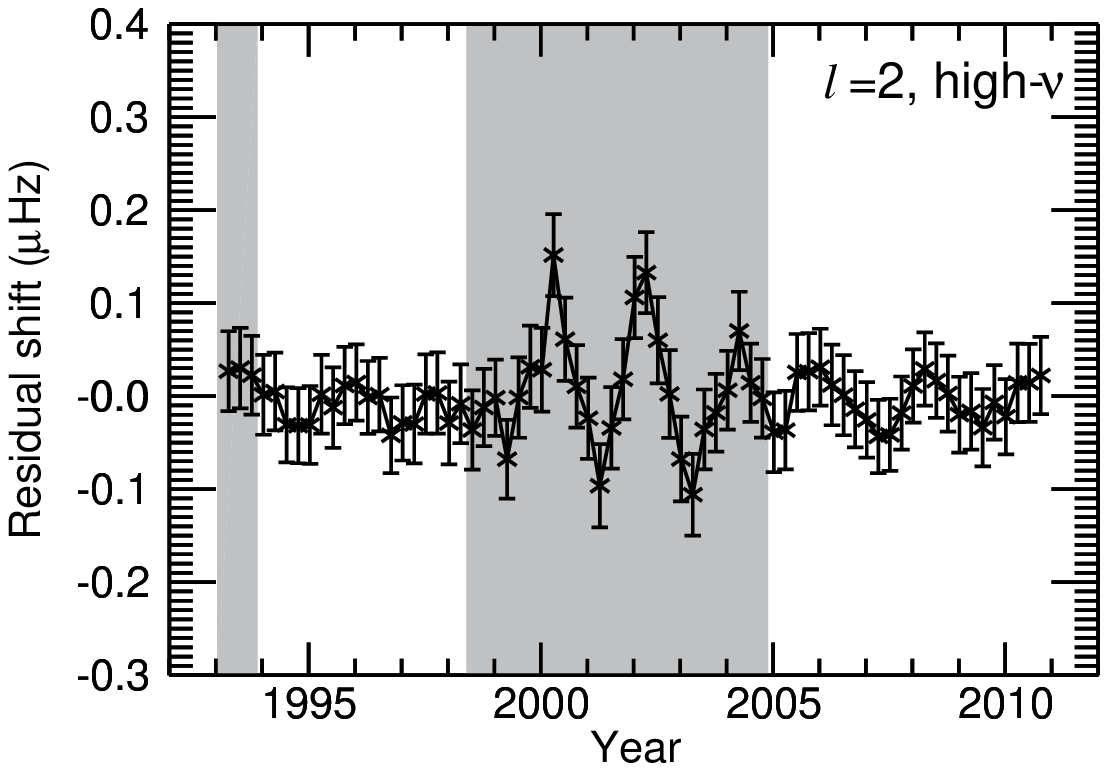}\\
  \vspace{0.5cm}
  \includegraphics[width=0.33\textwidth,
  clip]{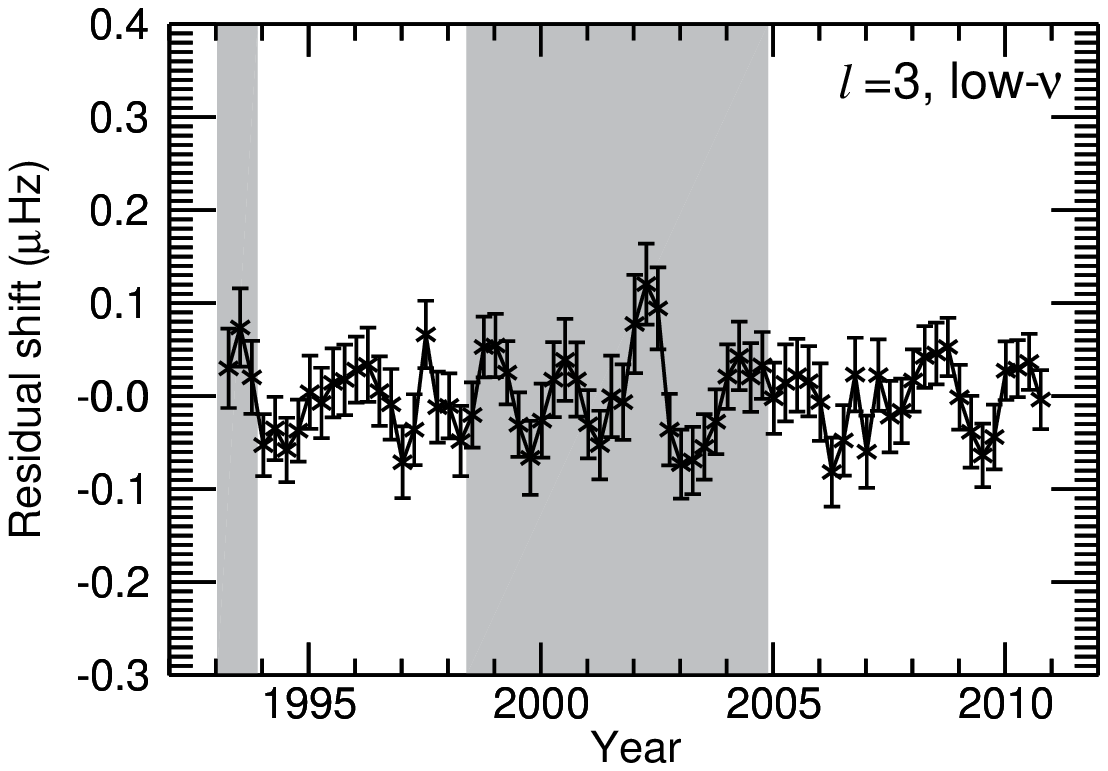}
  \includegraphics[width=0.33\textwidth,
  clip]{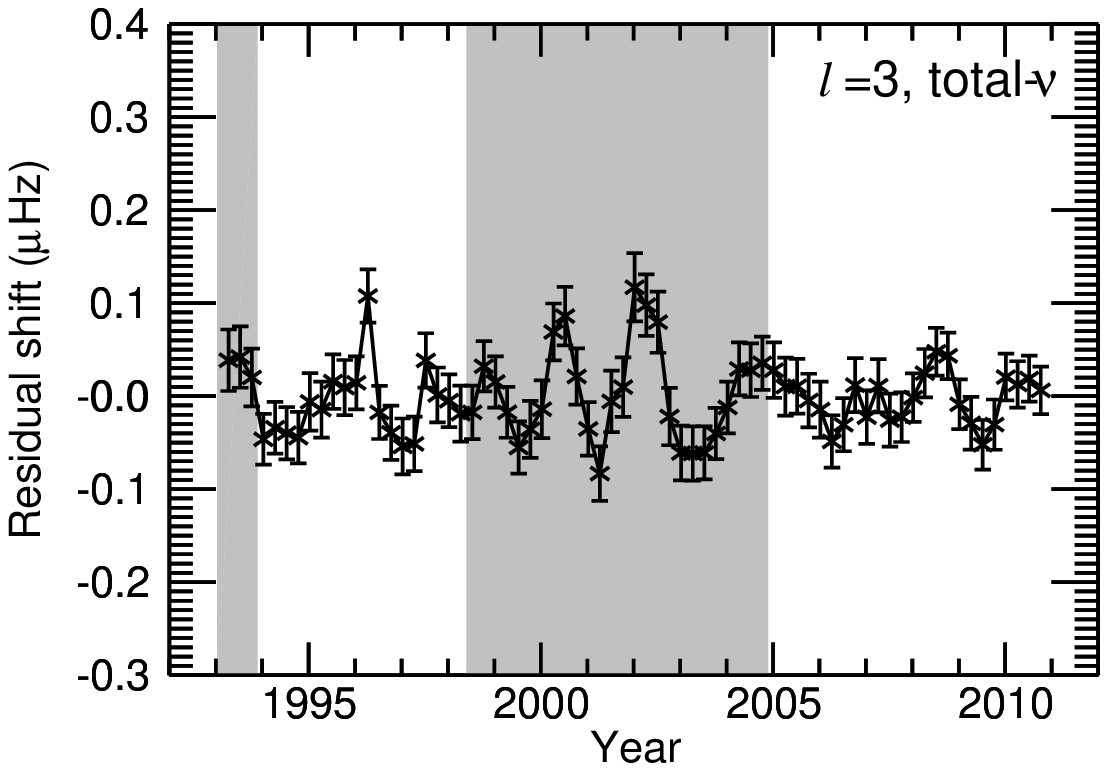}
  \includegraphics[width=0.33\textwidth, clip]{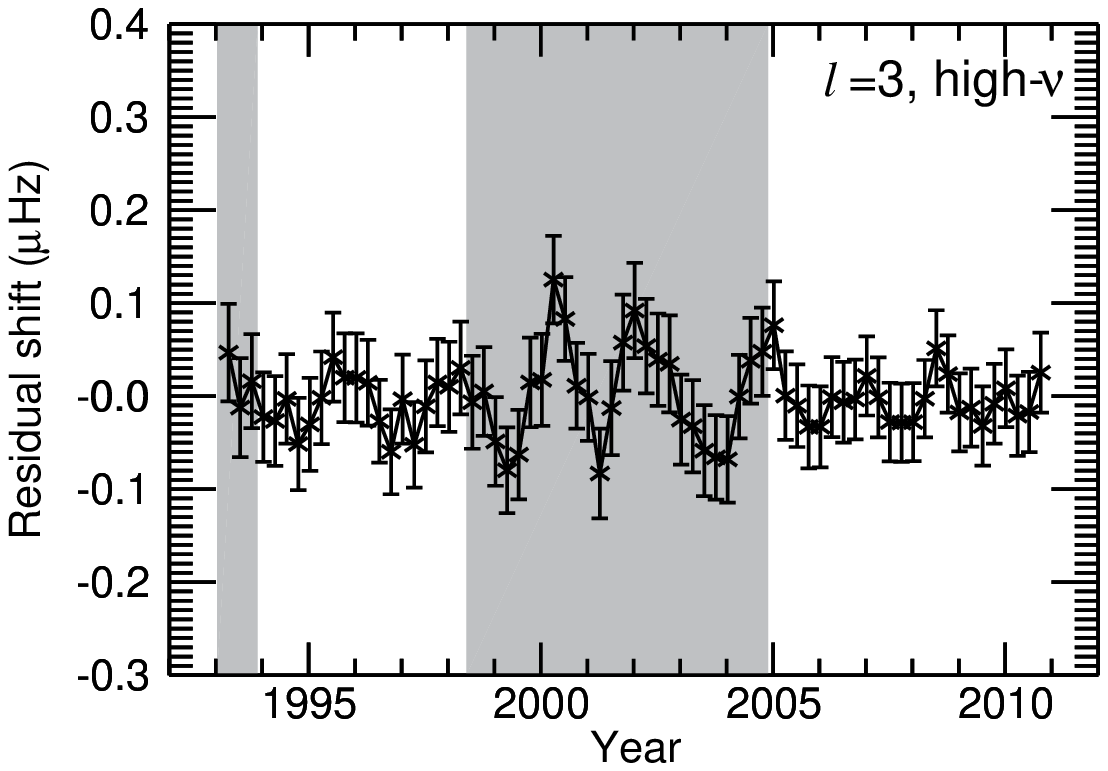}\\
  \caption{Individual-$l$ residuals left after the dominant 11-yr signal has been
  removed. The plotted results were obtained from the 365\,d subsets
  that overlapped by 91.25\,d.
  Right column: residuals for modes with frequencies between 1.86 and $2.77\,\rm mHz$ (low-frequency range).
  Middle column: residuals for modes with frequencies between 1.86 and $3.72\,\rm
  mHz$ (total-frequency range). Left column: residuals for modes with frequencies
  between 2.80 and $3.72\,\rm mHz$ (high-frequency range).
  First row: residuals for $l=0$ modes. Second row: residuals for
  $l=1$ modes. Third row: residuals for $l=2$ modes. Fourth row:
  residuals for $l=3$ modes. In each panel the shaded regions
  denote times of ``high-surface activity``,
  while the unshaded regions denote times of
  ``low-surface activity``. }\label{figure[residuals]}
\end{figure*}

We now look at the $l$-dependence of the QBO in more detail. In order to extract the seismic QBO, we subtracted a smooth trend
from the average shifts by applying a boxcar filter of width
2.5\,yr. This removed the dominant 11-yr signal of the solar cycle.
Note that, although the width of this boxcar is only slightly longer
than the periodicity we are examining, wider filters produce similar
results. The resulting residuals, which can be seen in Fig.
\ref{figure[residuals]} (from the overlapping 365-d subsets), show a periodicity on a timescale of about
2\,yrs i.e. the seismic QBO.

For each $l$, the amplitude of the seismic QBO in the total- and
high-frequency ranges varies with time. However, we note this is
only marginally true for the $l=1$ total-frequency range.
Furthermore, the amplitude of the signal in the $l=1$
total-frequency range appears to be smaller than the amplitude of
the signal in the $l=1$ low- and high-frequency range because, at
times, the signal in the low- and high-frequency ranges
destructively interfere with each other. It could be argued that the
amplitude of the seismic QBO in the total-frequency range is also
smaller than in the low- and high-frequency ranges for the other
$l$, although to a lesser extent. The effect is particularly evident
at solar minimum.

The amplitude of the seismic QBO is larger at times of high solar
surface activity than at times of low surface activity i.e. there is
evidence for an 11-yr envelope on top of the seismic QBO, which is
in agreement with the results of \citet{Fletcher2010}. The amplitude
of the envelope shows some degree dependence, being particularly
visible in the $l=2$ and 3 high-frequency range residuals and less
visible in the $l=0$ and 1 residuals. Notice that, for each $l$, the
same envelope is barely, if at all, visible in the
low-frequency-range residuals (see the left-hand panels of Fig.
\ref{figure[residuals]}) i.e. the amplitude of the seismic QBO
remains approximately constant with time. This indicates that the
effect of the 11-yr cycle on the seismic QBO is, to some extent,
frequency dependent.

The effect of the 11-yr envelope is further demonstrated by Table
\ref{table[abs dev]}, which shows the maximum absolute deviations of
the frequency residuals. For the high-frequency-range $l=2$ and 3
modes the maximum absolute deviation is greater at times of high
activity than at times of low activity. However, the maximum
absolute deviations observed in the high-frequency-range $l=0$ and 1
modes are approximately constant.

\begin{figure}
  \centering
  \includegraphics[width=0.48\textwidth,
  clip]{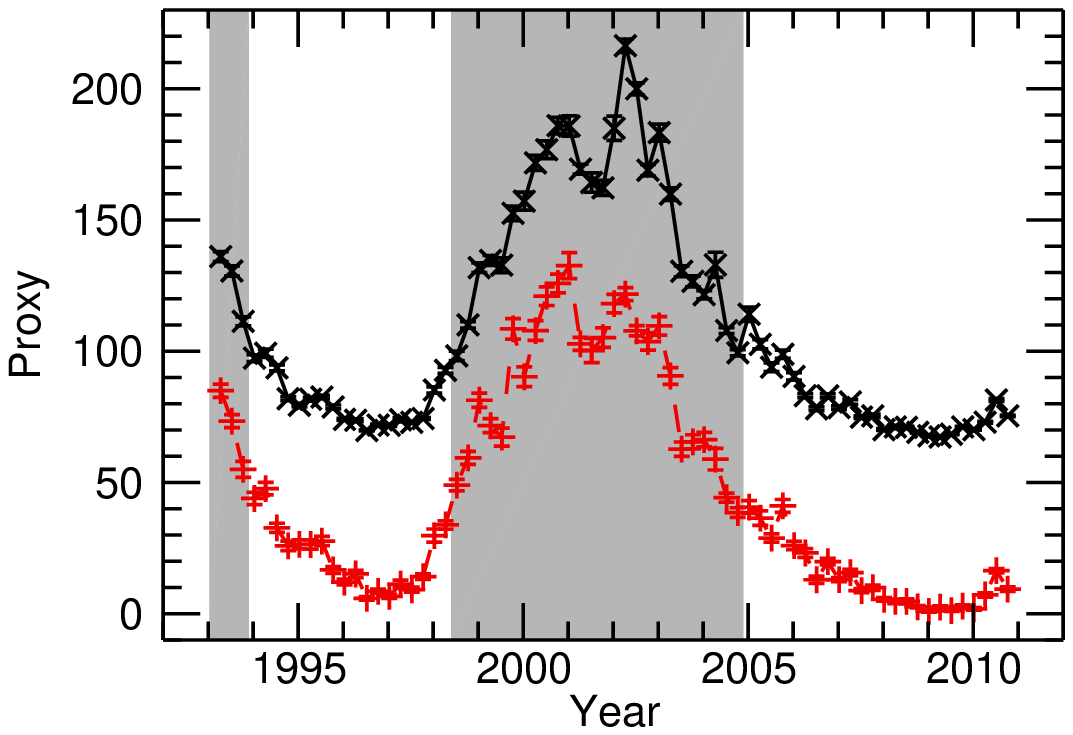}\\
  \includegraphics[width=0.48\textwidth,
  clip]{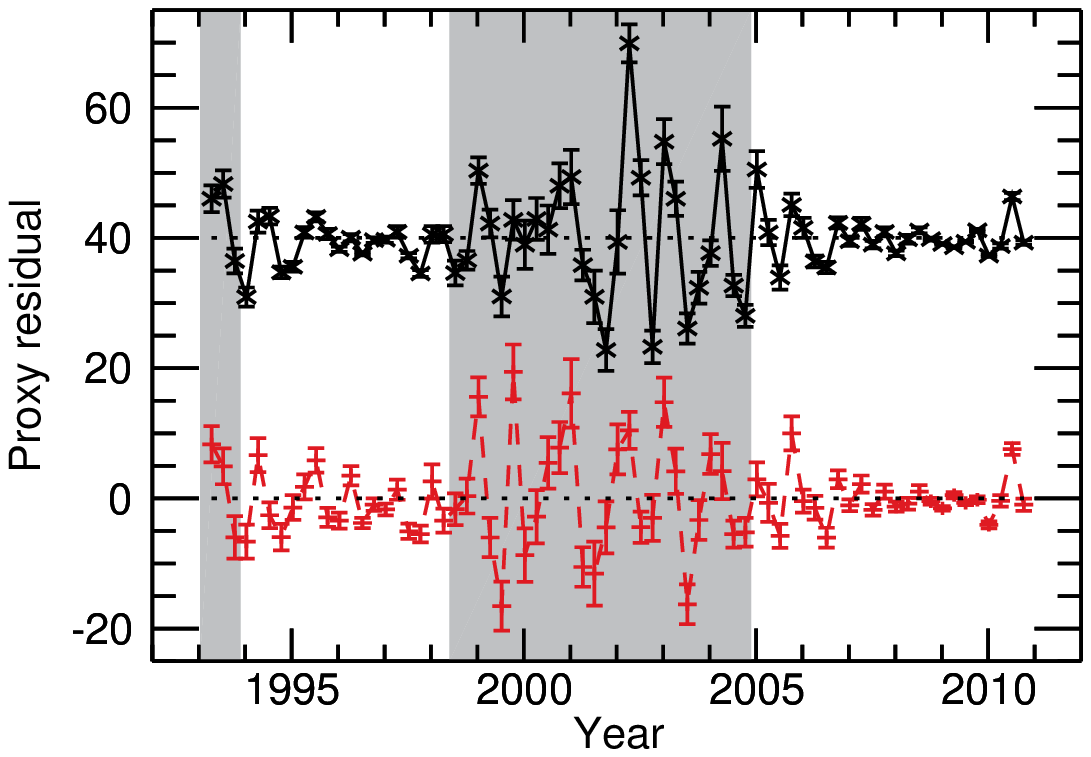}\\
  \caption{Top panel: Weighted mean activity proxies observed for
  365\,d overlapping subseries. The black solid line with the
  cross symbols shows the $\textrm{F}_{10.7}$, which is measured in
  units of $10^{-22}\,\rm W\,m^{-2}\,Hz^{-1}$. The red dashed line
  with the plus symbols shows the ISN. The errors on the proxies are determined using
  the standard error associated with a mean.
  Bottom panel: Residuals of the $\textrm{F}_{10.7}$ and the ISN after the
  dominant 11-yr signal was removed. The proxy residuals were found using
  365\,d overlapping subsets. The lines and symbols are the
  same as in the top panel. The $\textrm{F}_{10.7}$ residuals have been offset by $40\times10^{-22}\,\rm W\,m^{-2}\,Hz^{-1}$ to ease
  viewing.}\label{figure[proxies]}
\end{figure}

A similar dependence on the 11-yr solar cycle is also evident in the
QBO in other proxies of the Sun's activity \citep[e.g.][and
references therein]{Benevolenskaya1995, Hathaway2010, Mursula2003,
Valdes-Galicia2008, Vecchio2008, Vecchio2009, Zaqarashvili2010}. The
top panel of Fig. \ref{figure[proxies]} shows the variation
throughout the last solar cycle of the 10.7-cm radio flux
($\textrm{F}_{10.7}$) emitted from the Sun and the International
Sunspot number (ISN). Many authors \citep[e.g.][and references
therein]{Elsworth1990, Chaplin2007, Howe2008} have determined and commented upon
the good correlations between shifts in the p-mode frequencies and
activity proxies such as the $\textrm{F}_{10.7}$ and ISN. The bottom
panel of Fig. \ref{figure[proxies]} shows the residuals of the
proxies once the dominant 11-yr signal has been removed (obtained by
subtracting a smooth trend from the proxies). The proxies show clear
evidence of the QBO, predominantly at times of high solar
activity.

It is, therefore, reasonable to ask: Is there any evidence for the
seismic QBO during times of low activity? Although not shown here,
periodograms of the high-frequency-range frequency shifts observed
during the most recent solar minimum (the unshaded region after
2005 in Fig. \ref{figure[freq shifts]}) indicate that that seismic
QBO is still present, with a false alarm significance of less than
10\,per cent in the $l=0$, 1, and 2 p modes. The seismic QBO is not
significant at a 10\,per cent threshold level in the $l=3$ modes
because the errors associated with the frequency shifts are larger.

These results are not inconsistent with the conjecture that the magnetic flux responsible for the
seismic QBO is situated in the near surface shear layer. Under such a scenario, when the 11-yr cycle is in a
strong phase, buoyant magnetic flux sent upward from the base of the
envelope could help to nudge flux produced in the near surface shear layer into shallower regions thus allowing the seismic
QBO to be detected in the frequency signal. The 11-yr magnetic
activity which brings the flux responsible for the seismic QBO to the
surface is concentrated at low latitudes
($5^{\circ}\le\lambda\le40^{\circ}$). When the 11-yr cycle is at or
close to solar maximum, flux originating in the near-surface shear layer at low latitudes
will be closer to the surface than the flux at higher latitudes. The influence of the flux on the p-mode
frequencies is strongest when the flux is close to the surface. In
Sun-as-a-star data the frequencies determined for $l=2$ and $l=3$
modes are dominated by their sectoral components
\citep[$|m|=l$;][]{Chaplin2004b, Chaplin2004a}, which show a greater
sensitivity to low latitudes than the zonal components ($m=0$).

This conjecture is consistent with the observation that the seismic QBO
is stronger in the p modes and surface activity measures around
solar maximum. It is also understandable that the influence of the
11-yr signal on the flux responsible for the seismic QBO is more visible in
$l=2$ and 3 modes than in the $l=0$ and 1 modes.

\begin{table}\centering
\caption{Maximum absolute deviations in the residuals, observed in the 365\,d overlapping subsets, for the
  low- and high-frequency ranges
  (measured in $\rm\mu Hz$). The top half of the table shows
  the maximum absolute deviations observed during time of high solar
  activity i.e. before $\sim1995$ and between 1999-2005. The bottom half of the table gives
  the maximum absolute deviations observed during times of low
  solar activity i.e. between 1995-1999 and 2005-2011 (also see for example Fig.
  \ref{figure[residuals]}).}\label{table[abs dev]}
  \begin{tabular}{|c|c|c|c|c|}
    \hline
     & $l=0$ & $l=1$ & $l=2$ & $l=3$\\
    \hline
     & \multicolumn{4}{c}{During periods of high solar activity}\\
    \hline
    Low & $0.04\pm0.03$ & $0.04\pm0.02$ & $0.06\pm0.03$ & $0.12\pm0.03$ \\
    High & $0.07\pm0.03$ & $0.07\pm0.04$ & $0.14\pm0.04$ & $0.12\pm0.05$ \\
    \hline
     & \multicolumn{4}{c}{During periods of low solar activity}\\
    \hline
    Low & $0.03\pm0.03$ & $0.05\pm0.02$ & $0.06\pm0.03$ & $0.08\pm0.03$ \\
    High & $0.05\pm0.04$ & $0.07\pm0.03$ & $0.04\pm0.04$ & $0.08\pm0.05$ \\
    \hline
  \end{tabular}

\end{table}

\subsection{Constraints on the l-dependence of the seismic QBO}\label{section[QBO
constraints]} Monte Carlo simulations were performed to determine
how constrained the observed $l$-dependence of the seismic QBO is for the entire epoch examined in this paper. For each
degree 10,000 time series were simulated to contain a sine wave signal
with a frequency of $0.5\,\rm yr^{-1}$, and a power spectral
density comparable to that observed in the periodograms of the solar
minimum frequency shifts. Noise was added to the signal, using a
normally distributed random number generator, where the standard
deviation of each point in the time series was taken as the
$1\,\sigma$ error associated with each frequency shift plotted in
Fig. \ref{figure[freq shifts]}. The simulated time series were then
used to create periodograms. The simulations imply that there is
more than a 90\,per cent chance that the amplitude of the seismic
QBO in the $l=2$ and 3 modes is greater than in the $l=0$ modes.
Similarly there is more than an 80\,per cent chance that the
amplitude of the signal in the $l=2$ and 3 modes is greater than in
the $l=1$ modes. There is a 65\,per cent chance that the amplitude
of the signal in the $l=1$ modes is greater than in the $l=0$ modes,
while the amplitude of the signal in the $l=2$ and 3 modes is the same, to within the associated errors.

A similar analysis can be performed for the most recent solar
minimum. As the $l=3$ signal is not significant during the low
activity epoch we have only considered $0\le l\le 2$. The
simulations showed that the $l$-dependence of the seismic QBO is
poorly constrained by these observations and so it would be
difficult to make any inferences about the latitudinal distribution
of the source of the seismic QBO alone.

We also examined the $l$-dependence during the cycle 23 solar
maximum i.e. 1998-2005. There is more than a 99\,per cent
probability that the seismic QBO in the $l=2$ p modes is greater
than in both the $l=0$ and 1 p modes and there is more than a
95\,per cent chance that the seismic QBO in the $l=3$ modes is
greater than in the $l=0$ and 1 modes at solar maximum. The
amplitude of the signal in the $l=2$ and 3 modes is the same, to within the associated errors.

\begin{table}\centering
\caption{Ratios of the maximum absolute deviations of the $l=1$, 2,
and 3 modes with the maximum absolute deviations of the $l=0$ modes
for the high-frequency range residuals during periods of high solar
activity.}\label{table[ratios]}
  \begin{tabular}{|c|c|c|}
    \hline
    $l=1$ & $l=2$ & $l=3$\\
    \hline
    $1.0\pm0.7$ & $2.0\pm0.7$ & $1.7\pm0.8$ \\
    \hline
  \end{tabular}
\end{table}

Table \ref{table[ratios]} gives the ratios between the maximum
absolute deviations of the $l=1$, 2, and 3 modes and the maximum
absolute deviations of the $l=0$ modes for the high-frequency range
residuals during periods of high solar activity. Even at solar maximum
the constraints on the $l$-dependence of the seismic QBO are not
stringent and so it may be necessary to look at the frequency shifts
experienced by higher-$l$ modes before any definitive conclusions
can be made.

At solar maximum the seismic QBO appears to be modulated by the 11-yr solar
cycle, which is known to be strongest at low latitudes, and
therefore has more of an influence on the higher-$l$ modes.
Therefore, the $l$-dependence of the seismic QBO at solar maximum
contains information on the latitudinal dependence of both the 11-yr
solar cycle and the seismic QBO such that the latitudinal dependence of the seismic QBO alone cannot be extracted. To determine the latitudinal
dependence of the seismic QBO we must look at solar minimum.
However, as we have shown, at this time the signal is not strong
enough to constrain the $l$ dependence.

The amplitude of the QBO in the low-frequency-range $l=3$ residuals
is noticeably larger than for the other $l$ (see Fig.
\ref{figure[residuals]} and Table \ref{table[abs dev]}). This could
be a genuine effect: As noted in the previous section, there is a
well-known dependence on $l$ in the size of the 11-yr solar cycle
frequency shifts. However, we note that the estimates of the $l=3$
mode frequencies are more noisy than the frequencies estimated for
the other $l$. Furthermore,
estimates of the $l=3$ mode frequencies are also influenced by the
nearby, much stronger, $l=1$ modes. This is partially reflected in the size of
the error bars associated with the fitted frequencies (see also
Figs. \ref{figure[freq shifts]} and \ref{figure[residuals]}), but, as seen in Section \ref{subsection[significance of QBO]} it is still possible that the uncertainties are underestimated.

\subsection{The frequency dependence of the seismic
QBO}\label{subsection[freq dep qbo]} Fig. \ref{figure[residuals]}
shows that the seismic QBO is more evident at high frequencies and
it could be argued that the ``signal'' in the low-frequency range is
just noise. This implies that the seismic QBO signal shows some
frequency dependence.

We have used the 182.5\,d frequencies to determine the frequency at
which the seismic QBO stops being significant. To do this the
weighted average frequency shifts were generated for each subset in
time, in the manner described in Section \ref{section[data]}.
However, here we averaged the frequency shifts of the modes over
four overtones only. The lowest frequency range for which the mean shifts
were calculated was 1.86\,-\,2.37\,mHz, i.e. the lower limit of the
frequency ranges described in Section \ref{section[data]}. The next
frequency range was positioned so that it overlapped this range by 3
overtones i.e. 1.99\,-\,2.50\,mHz. This process was repeated until the
upper limit on the frequency ranges described in Section
\ref{section[data]} was reached. In total the mean
frequency shift was determined for 11 frequency ranges. We have
considered the $l=2$ frequency shifts only, because over the entire epoch
considered here, the seismic QBO is strongest in the $l=2$
frequencies (see Fig. \ref{figure[periodograms]}). A
periodogram of the $l=2$ frequency shifts was then determined in the
manner described in Section \ref{subsection[significance of QBO]}.
We found that the lowest frequency band at which the seismic QBO was
still significant at a 1\,per cent false alarm level was
2.60-3.04\,mHz.

\begin{figure}
\centering
  \includegraphics[width=0.4\textwidth, clip]{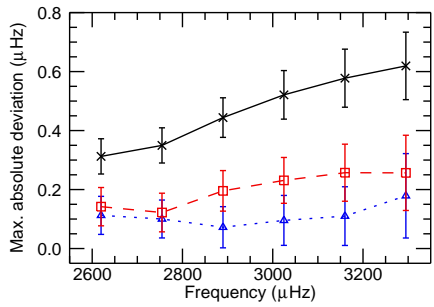}\\
  \caption{Maximum absolute deviations of $l=2$ residuals observed in
  different frequency bands. Each frequency band covers 4 overtones
  and the data are plotted at the frequency of the lowest overtone in each range. The
  red squares and the dashed line show the results for times of high-surface activity
  and the blue triangles and the dotted line show the results for times of low-surface
  activity. For comparison purposes the black crosses and the solid line show the
  maximum absolute deviations observed in the raw frequency shifts,
  which represents the amplitude of the seismic 11-yr solar cycle.}
  \label{figure[frequency dependence]}
\end{figure}

Residuals of the $l=2$ frequency shifts were then determined for the
frequency bands in which the seismic QBO was significant. The
maximum absolute deviations of the residuals at times of high- and
low-surface activity are shown in Fig. \ref{figure[frequency
dependence]}. For comparison purposes the maximum absolute deviation
of the raw frequency shifts was also calculated, as this reflects
the amplitude of the seismic 11-yr solar cycle. Fig.
\ref{figure[frequency dependence]} shows that the frequency
dependence of the seismic QBO is weaker than the frequency
dependence of the 11-yr solar cycle.

Although the 11-yr solar cycle magnetic flux is believed to be
generated at the base of the convection zone the main influence of
the 11-yr signal on the p-mode frequencies occurs in the upper few
100\,km of the convection zone. This is above the upper turning point
of the lowest frequency modes examined here and so explains the
observed frequency dependence of the 11-yr cycle.
\citet{Fletcher2010} postulated that the magnetic activity
responsible for the seismic QBO is positioned deeper below the
surface than the regions responsible for the 11-yr signal and as such the effect of
the QBO on the p-modes shows a weaker frequency dependence.

Fig. \ref{figure[frequency dependence]} also shows that the
frequency dependence of the seismic QBO is perhaps stronger at times of
high-surface activity than at times of low-surface activity, which
reflects the frequency dependence of the 11-yr envelope and again supports the idea that the 11-yr signal influences the QBO.

\section{Can the degree dependence be modeled in terms of
spatial distribution of magnetic field?}\label{section[models]}

We now attempt to characterise the latitudinal distribution of the
seismic QBO using the $l$ dependence of the signal described in
Section \ref{section[QBS]}, together with some simple models.
\citet{Chaplin2007a} show that inferences about the spatial
distribution of the 11-yr solar cycle can be made using simple
models of how the latitudinal distribution of solar activity affects
modes of different $l$ and $m$. \citet{Chaplin2011} used this technique to make reasonably accurate inferences concerning the maximum latitude at which the strong field from the 11-yr solar cycle is observed. Here we determine whether similar
models can be used to obtain information about the spatial
distribution of the source of the seismic QBO.

We have produced a simple series of models to describe the
latitudinal dependence of the magnetic activity responsible for the
seismic QBO. The models are based on those used by
\citet{Chaplin2007a}. We took the amplitude of the magnetic activity
to be uniform between some lower and upper bounds in latitude,
$\lambda$, and null everywhere else. We neglect any radial
dependence of the sound speed to simplify the problem. We assumed
that the expected frequency shifts caused by the modeled magnetic
activity were proportional to \citep[e.g.][]{Moreno2000}
\begin{equation}\label{equation[shift]}
    \delta\nu_{lm}\propto\left(l+\frac{1}{2}\right)\frac{\left(l-m\right)!}
    {\left(l+m\right)!}\int_0^\pi|P_l^m(\cos\theta)|^2b(\theta)\sin\theta\textrm{d}\theta,
\end{equation}
where $P_l^m(\cos\theta)$ are Legendre functions, while
$b(\theta)$ describes the magnetic activity and, more specifically,
its distribution in colatitude, $\theta=90-\lambda$. We assumed that
when the frequency residuals were zero the activity responsible for
the seismic QBO was zero at all latitudes, and consequently
$b(\theta)=0$ for all $\theta$. When, $b(\theta)\ne0$ for a chosen
range of $\theta$, the determined $\delta\nu_{lm}$ then correspond
to the amplitude of the seismic QBO.

For the chosen range of $\theta$ the magnetic flux was set to some constant value $(b(\theta)=\textrm{constant}\ne0)$. The
only parameters that were changed from model to model were the
maximum and minimum latitudes at which activity was present,
$\lambda_{\textrm{\scriptsize{max}}}$ and
$\lambda_{\textrm{\scriptsize{min}}}$ respectively. For each
combination of $\lambda_{\textrm{\scriptsize{max}}}$ and
$\lambda_{\textrm{\scriptsize{min}}}$ we calculated a value for
$\delta\nu_{lm}$.  To summarise:
\begin{equation}\label{equation[b model 1]}
    b(\theta)=\left\{
\begin{array}{cl}
  \textrm{constant,} & \textrm{for }\lambda_{\textrm{\scriptsize{min}}}\le|\lambda|\le\lambda_{\textrm{\scriptsize{max}}}; \\
  0, & \textrm{otherwise.}\\
\end{array}
\right.
\end{equation}
Note that, in the above scenario the total magnetic flux in the
region of influence is proportional to the area of the Sun's surface
between $\lambda_{\textrm{\scriptsize{max}}}$ and
$\lambda_{\textrm{\scriptsize{min}}}$ and therefore varied
considerably over the range of computations made.

We have made no restricting assumptions over the latitudinal range over which the
magnetic activity responsible for the QBO is present. So for each
$\lambda_{\textrm{\scriptsize{min}}}$ in the range
$0^{\circ}\le\lambda_{\textrm{\scriptsize{min}}}<90^{\circ}$ we
determined the frequency shift when
$\lambda_{\textrm{\scriptsize{max}}}$ was varied between
$\lambda_{\textrm{\scriptsize{min}}}<\lambda_{\textrm{\scriptsize{max}}}\le90^{\circ}$.
The aim of this Section is then to find a combination of values for
$\lambda_{\textrm{\scriptsize{max}}}$ and
$\lambda_{\textrm{\scriptsize{min}}}$ that best represents the
observed results.

The results in Fig. \ref{figure[residuals]} show that the residuals
are modulated by an 11-yr solar cycle envelope. However, according
to our earlier explanation of the seismic QBO, the amplitude of the
signal is enhanced at times of high activity by the 11-yr cycle and the magnitude of the enhancement is dependent on $l$.
Therefore, the $l$-dependence of the amplitude of the QBO occurs because of the spatial distribution of the 11-yr solar cycle rather than the flux responsible for the seismic QBO. However, during epochs of low surface activity
the signal from the seismic QBO is not strong enough to allow
constraints to be put on the $l$-dependence of the signal.
Therefore, we use the ratios from times of high surface activity
that are shown in Table \ref{table[ratios]} to constrain the models.
We also note that the Monte Carlo simulations performed in Section
\ref{section[QBO constraints]} imply that at solar maximum the
seismic QBO in the $l=2$ and 3 modes will be larger than in the
$l=1$ modes.

\begin{figure*}
  \includegraphics[width=0.4\textwidth, clip]{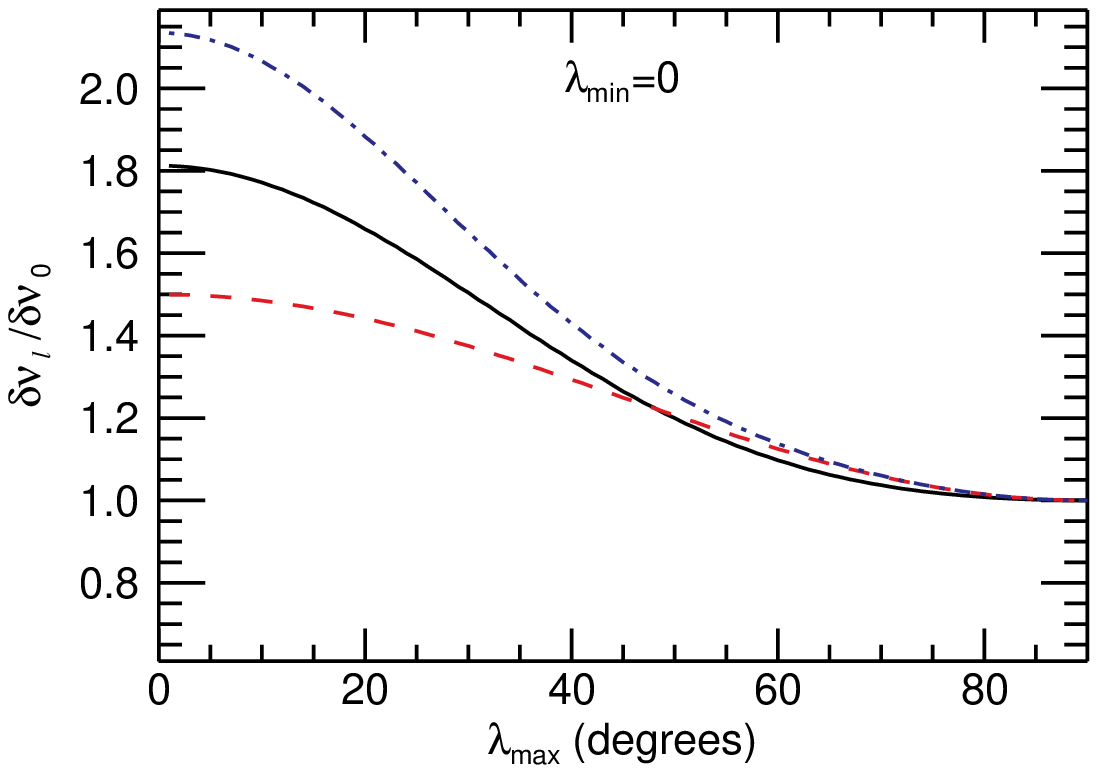}
  \includegraphics[width=0.4\textwidth, clip]{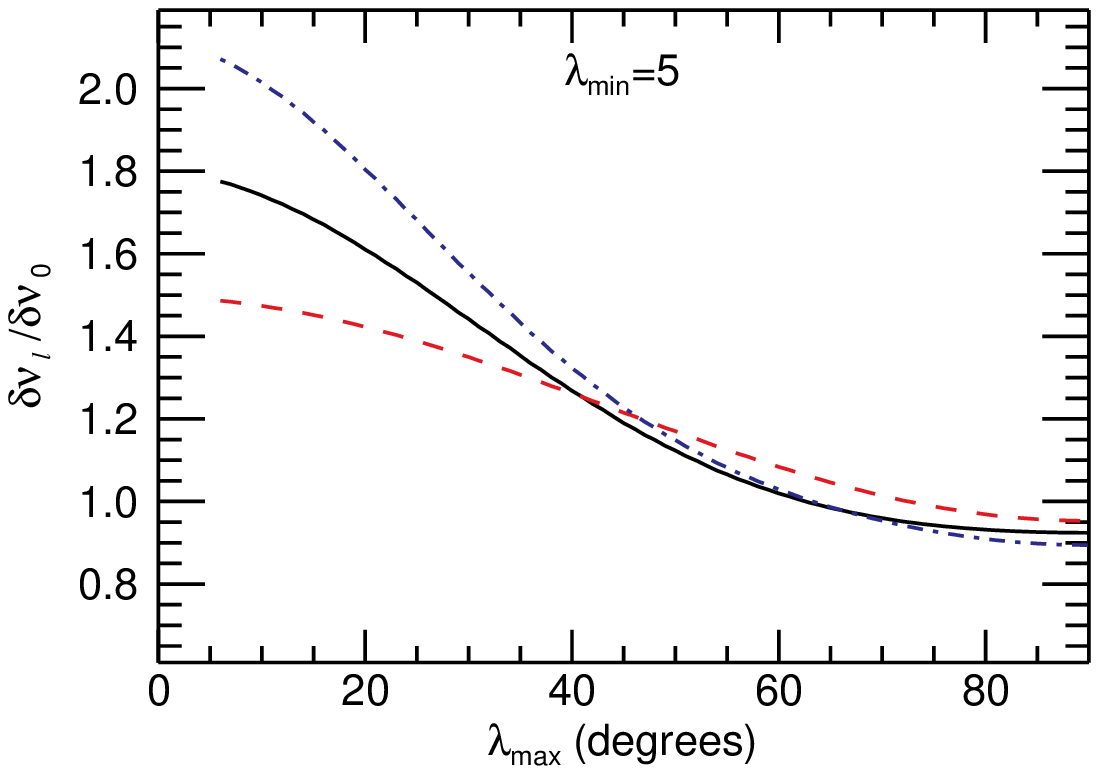}\\
  \includegraphics[width=0.4\textwidth, clip]{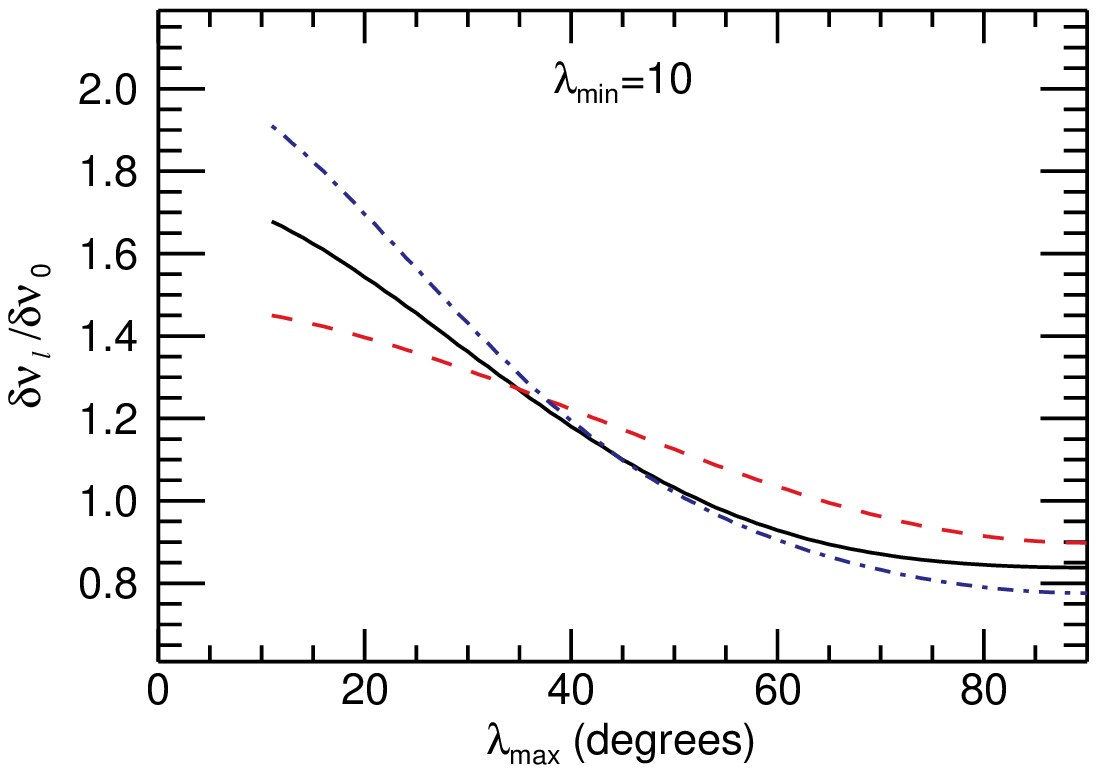}
  \includegraphics[width=0.4\textwidth, clip]{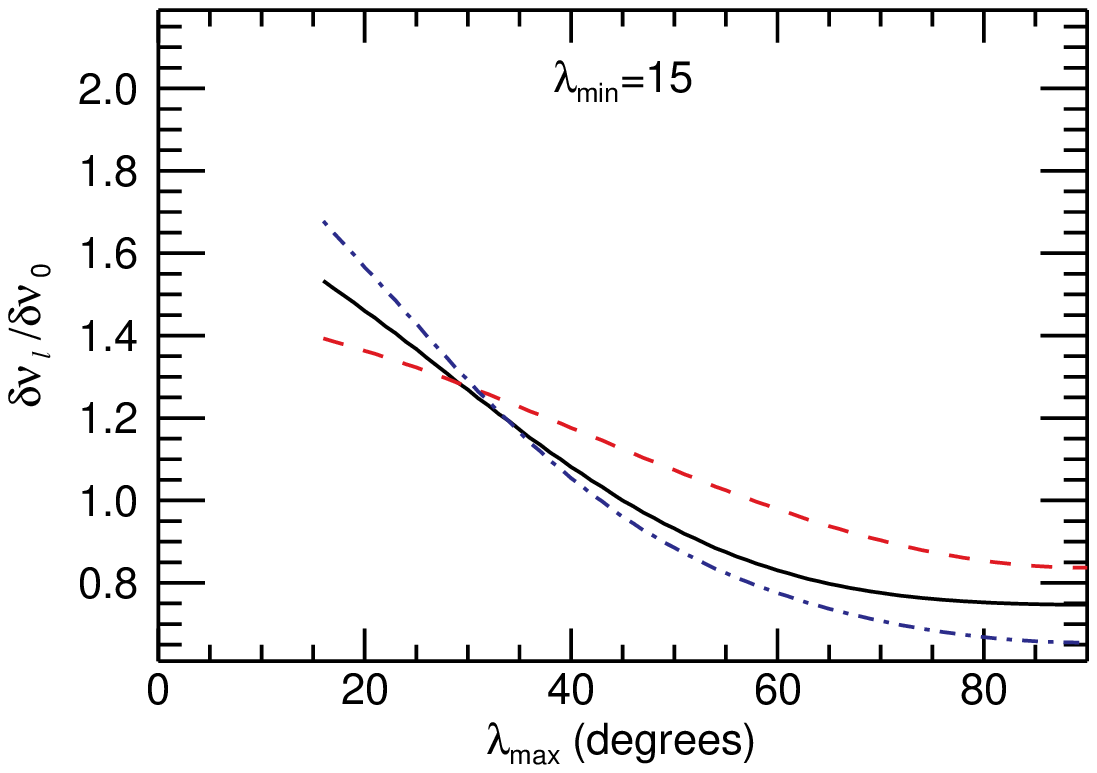}\\
  \caption{Ratios of the $l=1$, 2, and 3 model frequency shifts and the $l=0$ model frequency shifts given by simple models of magnetic
  activity. In each panel the linestyles are as
  follows: the red dashed line represents
  $l=1$; the black solid line represents $l=2$; and the
  blue dot-dashed line represents $l=3$.}\label{figure[model]}
\end{figure*}

In Sun-as-a-star data only modes where $l+m$ is even are visible,
because of the near-perpendicular inclination of the solar rotation
axis with respect to the Earth. When the frequency of an $l=2$ mode
is obtained from Sun-as-a-star data, 90\,per cent of the final
frequency is determined by the $|m|=2$ components and only 10\,per
cent is determined by the $m=0$ component. Similarly when estimating
$l=3$ frequencies 94\,per cent of the final frequency is determined
by the $|m|=3$ components and only 6\,per cent by the $|m|=1$
components \citep{Chaplin2004b, Chaplin2004a}. The $l=2$ and 3 model
frequencies shifts were, therefore, constructed taking these
percentages into account.

Ratios between the $l=1$, 2, and 3 model frequency shifts and the
$l=0$ model frequency shifts were computed i.e.
$(\delta\nu_l/\delta\nu_0)$ and the results of a few examples of the
model series are shown in Fig. \ref{figure[model]}. The top left-hand panel of Fig.
\ref{figure[model]} shows that when
$\lambda_{\textrm{\scriptsize{min}}}=0^\circ$ the constraints on the
ratios are satisfied when
$\lambda_{\textrm{\scriptsize{max}}}<45^\circ$. As
$\lambda_{\textrm{\scriptsize{min}}}$ increases the
$\lambda_{\textrm{\scriptsize{max}}}$ at which the constraints are
satisfied decreases. The models, therefore, imply that the QBO is
strongest in equatorial regions. This is to be expected if the QBO
is indeed modulated by the buoyant strong flux from the 11-yr solar cycle,
which is mainly observed in equatorial regions.

\section{Summary}\label{section[discussion]}

We have shown that the seismic QBO is present in the individual-$l$
frequency shifts. The main properties of the seismic QBO are:
\begin{enumerate}
    \item The seismic QBO in the p-mode frequencies shows
    an 11-yr envelope. The envelope is most visible in the $l=2$ and 3 modes but is also present in the
    $l=0$ and 1 p-mode frequencies. A possible explanation for the difference in the amplitude of the envelope is that close to
    solar maximum, the buoyant strong flux from the 11-yr
    cycle brings the flux responsible for the seismic QBO to the surface, where its influence on the p-mode frequencies is stronger.
    The buoyant strong flux from the 11-yr cycle is mainly
    observed at low latitudes, which the $l=2$ and 3 modes are more
    sensitive to than the $l=0$ and 1 modes. The $l$-dependence of the
    11-yr envelope would, therefore, be influenced by the 11-yr cycle and
    so would not tell us about the latitudinal distribution of the
    seismic QBO alone.
    \item To determine the latitudinal distribution of the flux responsible for the seismic QBO we
    must study the seismic QBO signal away from solar maximum.
    Although significant at a 10\,per cent false alarm level, the amplitude of
    the seismic QBO at times of solar minimum is much smaller than
    at solar maximum. Therefore it was not possible to constrain the latitudinal dependence of the QBO alone.
    \item We have used simple models to determine the latitudinal distribution of the flux
    responsible for the seismic QBO when the 11-yr cycle is at its maximum. The models imply that
    the $l$-dependence of the seismic QBO is best replicated when
    the magnetic flux is between $\pm45^\circ$. This would be expected if,
    at solar maximum, the seismic QBO is modulated by the 11-yr
    cycle and the strong-field component of the 11-yr cycle is mainly observed at
    low-latitudes.
    \item The seismic QBO is dependent on mode frequency but this
    frequency dependence is weaker than in the frequency shifts of the 11-yr solar cycle. The
    frequency dependence of the seismic QBO is stronger at solar
    maximum than at solar minimum. This result is consistent with the conjecture that the flux responsible for the seismic QBO is dragged to the
    surface by the buoyant magnetic flux of the 11-yr solar cycle at
    solar maximum.

\end{enumerate}

In terms of the latitudinal distribution of the flux responsible for
the seismic QBO, the results of this paper are by no means
conclusive. To determine whether the latitudinal dependence of the
seismic QBO agrees with the latitudinal distribution proposed by
\citet{Vecchio2009}, where the signal is strong in polar regions and
weaker around the equator, we would need to consider times of low
surface activity. Extending this study to higher-$l$, where it is
possible to study a wider range of $m$ components, may help resolve
this matter. However, it is also possible that, as with the low-$l$ modes, the latitudinal dependence of the seismic QBO cannot be isolated from the latitudinal dependence of the 11-yr solar cycle. Observing the Sun throughout the next solar cycle and
beyond will enable better comparisons between the different $l$ to
be made. The recent solar minimum has been described as unusual
because of its depth and longevity. It will be interesting,
therefore, to compare the seismic QBO observed in the upcoming solar
cycle with that observed in the last.

\section*{Acknowledgements}

This paper utilizes data collected by the Birmingham
Solar-Oscillations Network (BiSON). We thank the members of the
BiSON team, both past and present, for their technical and
analytical support. We also thank P. Whitelock and P. Fourie at
SAAO, the Carnegie Institution of Washington, the Australia
Telescope National Facility (CSIRO), E.J. Rhodes (Mt. Wilson,
Californa) and members (past and present) of the IAC, Tenerife.
BiSON is funded by the Science and Technology Facilities Council
(STFC). AMB, WJC, and YE also acknowledge the financial support of STFC. We thank R. New, D.W. Hughes and S.M. Tobias for helpful comments and discussions. RS
is grateful to S. Turck-Chi{\'{e}}ze and CEA/DAPNIA for providing the
facilities required to continue her work.

\bibliographystyle{mn2e_new}
\bibliography{l_dependence}
\end{document}